\documentclass[hidelinks,preprint,12pt]{elsarticle}
\usepackage{amsmath}
\usepackage{booktabs}
\usepackage{tikz}
\usepackage{graphicx}
\usepackage{pdflscape}
\UseRawInputEncoding

\usepackage{caption}

\tikzset{
    datanode/.style={rectangle, draw, rounded corners, fill=gray!15},
    processnode/.style={rectangle, draw, fill=blue!10},
    apinode/.style={rectangle, draw, fill=orange!20},
    llmnode/.style={rectangle, draw, fill=red!15},
    enrichnode/.style={rectangle, draw, fill=yellow!30},
    extendnode/.style={rectangle, draw, fill=green!20}
}

\usepackage{graphicx} 
\DeclareGraphicsExtensions{.pdf,.png}
\usepackage{hyperref} 
\usepackage[inline]{enumitem} 
\usepackage{array}
\usepackage{longtable}
\usepackage{booktabs}
\usepackage{tabularx}
\usepackage{pgfplots}
\pgfplotsset{compat=1.18}
\usepackage{subcaption}
\usepackage{multirow} 
\usepackage{lipsum} 
\usepackage{soul}
\usepackage{todonotes}
\usepackage{amssymb}
\usepackage{orcidlink}
\usepackage{hyperref}
\usepackage{verbatim}
\usepackage{amsmath}
\usepackage{listings}
\usepackage{xcolor}
\usepackage{booktabs}
\usepackage{siunitx}
\usepackage{multirow}
\usepackage{siunitx}
\usepackage{float}
\usepackage{threeparttable}
\usepackage{rotating}

\usepackage{soul}
\usepackage{xcolor}
\usepackage{cleveref}
\usepackage{enumitem}

\definecolor{highlightpurple}{RGB}{186, 85, 211} 

\definecolor{DeepSkyBlue}{RGB}{0, 191, 255}
\definecolor{ForestGreen}{RGB}{34, 139, 34}

\journal{Information and Software Technology}

\begin{document}

\begin{frontmatter}



\lstdefinelanguage{prompt}{
  basicstyle=\ttfamily\small\color{black},
  commentstyle=\color{gray}\itshape,
  stringstyle=\color{ForestGreen},
  morecomment=[l]{\#},
  morestring=[b]",
  sensitive=true,
  morekeywords={},
  keywordstyle=\color{DeepSkyBlue},
  backgroundcolor=\color{yellow!5},
  breaklines=true,
  frame=single,
  framesep=5pt,
  keepspaces=true,
  numbers=none,
  showspaces=false,
  showstringspaces=false,
  showtabs=false,
  captionpos=b,
  xleftmargin=15pt,
  xrightmargin=15pt,
  escapeinside={(*}{*)},
}

\lstset{
  language=prompt,
  basicstyle=\small\ttfamily,
  backgroundcolor=\color{gray!10},
  breaklines=true,
  captionpos=b,
  keepspaces=true,
  numbers=none,
  numbersep=5pt,
  showspaces=false,
  showstringspaces=false,
  showtabs=false,
  tabsize=2,
  frame=single,
  framesep=5pt,
  framerule=0.4pt,
  rulecolor=\color{black!30},
  xleftmargin=10pt,
  xrightmargin=10pt,
}
\lstdefinelanguage{text}{
  basicstyle=\ttfamily\scriptsize,
  commentstyle=\color{gray}\itshape,
  stringstyle=\color{ForestGreen},
  morestring=[b]",
  morecomment=[l]{\#},
  sensitive=true,
  morekeywords={},
  keywordstyle=\color{blue},
  backgroundcolor=\color{gray!5},
  breaklines=true,
  breakatwhitespace=true,
  frame=single,
  framesep=5pt,
  keepspaces=true,
  numbers=none,
  showspaces=false,
  showstringspaces=false,
  showtabs=false,
  tabsize=2,
  captionpos=b,
  xleftmargin=15pt,
  xrightmargin=15pt,
  escapeinside={(*}{*)}, 
}

\lstdefinelanguage{jsonschema}{
  morekeywords={true, false, null},
  sensitive=true,
  comment=[l]{//},
  morecomment=[s]{/*}{*/},
  commentstyle=\color{gray}\itshape,
  morestring=[b]{"}, 
  stringstyle=\color{ForestGreen},
  morekeywords=[2]{string, integer, float, boolean, file, directory, array, object},
  keywordstyle=[2]\color{violet},
  morekeywords=[3]{metadata, pipeline_summary, components, detailed_flow_structure, parameters, integrations},
  keywordstyle=[3]\color{blue}\bfseries,
  morekeywords=[4]{"id":, "name":, "type":, "description":, "inputs":, "outputs":, "image":, 
                   "is_internally_parallelized":, "timestamp":, "source_description_file":,
                   "llm_provider":, "llm_model_key":, "analysis_version":, "execution_environment":, 
                   "flow_pattern_summary":, "entry_points":, "nodes":, "next_nodes":, "parallel_blocks":, 
                   "triggered_by_nodes":, "instance_parameter":, "task_sequence_types":, "synchronization_node":,
                   "global":, "components":, "default":, "required":, "constraints":, "environment_variables":,
                   "integration_points":, "connection":, "authentication":, "direction":, "data_sources":,
                   "data_sinks":, "url":, "path":},
  keywordstyle=[4]\color{RoyalBlue},
  morekeywords=[5]{DataLoader, Transformer, Enricher, Exporter, QualityCheck, Splitter, Merger, Orchestrator},
  keywordstyle=[5]\color{teal},
  morekeywords=[6]{API, Database, FileSystem, MessageQueue, CloudStorage},
  keywordstyle=[6]\color{orange},
  morekeywords=[7]{input, output, both},
  keywordstyle=[7]\color{purple},
  backgroundcolor=\color{gray!5},
  basicstyle=\ttfamily\small,
  breaklines=true,
  breakatwhitespace=true,
  frame=single,
  framesep=5pt,
  keepspaces=true,
  numbers=left,
  numberstyle=\tiny\color{gray},
  numbersep=5pt,
  showspaces=false,
  showstringspaces=false,
  showtabs=false,
  tabsize=2,
  captionpos=b,
  xleftmargin=15pt,
  xrightmargin=15pt,
  escapeinside={(*}{*)}, 
}

\newcommand{\jsonschema}[2][]{%
  \lstset{language=jsonschema}%
  \lstinputlisting[#1]{#2}%
}

\lstnewenvironment{jsonschemalisting}[1][]
  {\lstset{language=jsonschema,#1}}
  {}

\lstdefinelanguage{json}{
  basicstyle=\ttfamily\scriptsize,
  commentstyle=\color{gray}\itshape,
  stringstyle=\color{ForestGreen},
  morestring=[b]",
  morecomment=[l]{//},
  morecomment=[s]{/*}{*/},
  literate=
    *{0}{{{\color{red}0}}}{1}%
    {1}{{{\color{red}1}}}{1}%
    {2}{{{\color{red}2}}}{1}%
    {3}{{{\color{red}3}}}{1}%
    {4}{{{\color{red}4}}}{1}%
    {5}{{{\color{red}5}}}{1}%
    {6}{{{\color{red}6}}}{1}%
    {7}{{{\color{red}7}}}{1}%
    {8}{{{\color{red}8}}}{1}%
    {9}{{{\color{red}9}}}{1}%
    {:}{{{\color{blue}{:}}}}{1}%
    {,}{{{\color{blue}{,}}}}{1}%
    {\{}{{{\color{blue}{\{}}}}{1}%
    {\}}{{{\color{blue}{\}}}}}{1}%
    {[}{{{\color{blue}{[}}}}{1}%
    {]}{{{\color{blue}{]}}}}{1},
  sensitive=true,
  morekeywords={true,false,null},
  keywordstyle=\color{blue},
}

\lstdefinelanguage{yaml}{
  basicstyle=\ttfamily\scriptsize,
  commentstyle=\color{gray}\itshape,
  stringstyle=\color{ForestGreen},
  morestring=[b]",
  morecomment=[l]{\#},
  sensitive=true,
  morekeywords={true,false,null},
  keywordstyle=\color{blue},
  backgroundcolor=\color{gray!5},
  breaklines=true,
  breakatwhitespace=true,
  frame=single,
  framesep=5pt,
  keepspaces=true,
  numbers=left,
  numberstyle=\tiny\color{gray},
  numbersep=5pt,
  showspaces=false,
  showstringspaces=false,
  showtabs=false,
  tabsize=2,
  captionpos=b,
  xleftmargin=15pt,
  xrightmargin=15pt,
  escapeinside={(*}{*)}, 
}

\lstdefinelanguage{python}{
  basicstyle=\ttfamily\scriptsize,
  commentstyle=\color{gray}\itshape,
  stringstyle=\color{ForestGreen},
  morestring=[b]",
  morestring=[b]',
  morecomment=[l]{\#},
  morekeywords={True,False,None,def,return,import,from,as,class,for,while,if,elif,else,try,except,finally,raise,with,lambda},
  keywordstyle=\color{blue},
  sensitive=true,
  backgroundcolor=\color{gray!5},
  breaklines=true,
  breakatwhitespace=true,
  frame=single,
  framesep=5pt,
  keepspaces=true,
  numbers=left,
  numberstyle=\tiny\color{gray},
  numbersep=5pt,
  showspaces=false,
  showstringspaces=false,
  showtabs=false,
  tabsize=2,
  captionpos=b,
  xleftmargin=15pt,
  xrightmargin=15pt,
  escapeinside={(*}{*)}, 
}

\newcommand{\breaktt}[1]{\texttt{\begingroup\obeyspaces\obeylines\breaktext#1\endgroup}}
\def\breaktext#1{\StrSubstitute{#1}{_}{\_\allowbreak}}

\title{Prompt2DAG: A Modular Methodology for LLM-Based Data Enrichment Pipeline Generation}

\author[1]{Abubakari Alidu\,\orcidlink{0009-0000-9251-919X}}
\author[1]{Michele Ciavotta\,\orcidlink{0000-0002-2480-966X}}
\author[1]{Flavio De~Paoli\,\orcidlink{0000-0002-5047-7371}}

\affiliation[1]{organization={Dept. Informatics, Systems and Communication, University of Milan-Bicocca},
                country={Italy}}

\begin{abstract}
Developing reliable data enrichment pipelines demands significant engineering expertise. We present Prompt2DAG, a methodology that transforms natural language descriptions into executable Apache Airflow DAGs. We evaluate four generation approaches—Direct, LLM-only, Hybrid, and Template-based—across 260 experiments using thirteen LLMs and five case studies to identify optimal strategies for production-grade automation. Performance is measured using a penalized scoring framework that combines reliability with code quality (SAT), structural integrity (DST), and executability (PCT). 

The Hybrid approach emerges as the optimal generative method, achieving a 78.5\% success rate with robust quality scores (SAT: 6.79, DST: 7.67, PCT: 7.76). This significantly outperforms the LLM-only (66.2\% success) and Direct (29.2\% success) methods. Our findings show that reliability, not intrinsic code quality, is the primary differentiator. Cost-effectiveness analysis reveals the Hybrid method is over twice as efficient as Direct prompting per successful DAG. We conclude that a structured, hybrid approach is essential for balancing flexibility and reliability in automated workflow generation, offering a viable path to democratize data pipeline development.
\end{abstract}




\end{frontmatter}

\section{Introduction}

Data pipelines are foundational to modern data-driven organizations, enabling the systematic collection, processing, transformation, and analysis of information across diverse domains~\cite{tiwari2023data, sresth2023optimizing}. These pipelines orchestrate workflows that integrate heterogeneous data sources, apply specific business logic, and deliver actionable insights. However, the design, implementation, and maintenance of such pipelines traditionally demand specialized data engineering expertise, creating a significant barrier for domain experts who possess subject-matter knowledge but may lack the specific programming skills needed to translate their requirements into robust, executable workflows~\cite{Dureja2024, Nara12134}.

This technical barrier restricts broader participation in data pipeline development, thereby limiting the agility with which organizations can innovate and respond to data-driven opportunities. The challenge is particularly evident when using powerful workflow orchestration frameworks that offer flexibility and comprehensive ecosystems, but require careful management of task dependencies, diligent configuration of operators, and often complex debugging processes~\cite{singh2019airflow}.

The capabilities demonstrated by Large Language Models (LLMs) in natural language understanding and code generation open up the possibility of bridging this expertise gap by enabling pipeline descriptions in natural language and translating them into structured workflow definitions or executable code~\cite{li2023structured, jiang2024survey}. However, the efficacy and reliability of LLM-based pipeline generation are heavily contingent on how the generation task is structured and how the LLM's capabilities are harnessed. Direct, end-to-end generation from high-level descriptions to complex code often struggles with precision, maintainability, and adherence to software engineering best practices~\cite{rao2025navigating}.

In this paper, we introduce and evaluate \textit{Prompt2DAG}, a methodology for transforming natural language descriptions into executable data workflows. Our work focuses specifically on data enrichment pipelines, a critical and broadly applicable category of data processing where organizations enhance their data assets through geocoding, entity reconciliation, or sentiment analysis. We systematically compare four distinct generation strategies—direct, LLM-only, hybrid, and template-based—to identify the architectural patterns that are essential for achieving production-grade reliability. Our primary contribution is the empirical demonstration that a hybrid, template-guided approach provides the optimal balance of flexibility and robustness, significantly outperforming purely generative methods.

Unlike end-to-end generation approaches, Prompt2DAG follows a structured, four-stage process-Pipeline Analysis, Workflow Description Generation, Executable Workflow Generation, and Automated Evaluation—to enhance clarity, facilitate validation at each stage, and make workflow creation more manageable and accessible, particularly for users who are not data engineering specialists.

Our experiments provide a comprehensive evaluation by leveraging a diverse suite of thirteen state-of-the-art LLMs, including both leading proprietary models like GPT-4o Mini and Claude 3.5 Sonnet, and a wide range of open-source architectures. This ensures our findings are representative of the current state of the art in language model capabilities.

The experimental results confirm that a structured, multi-stage approach is critical for reliable DAG generation. The deterministic Template-based method achieved the highest success rate (92.3\%), establishing a benchmark for reliability. Among generative methods, our Hybrid approach proved to be the most effective, delivering a robust 78.5\% success rate and the highest quality scores across all metrics (SAT: 6.79, DST: 7.67, PCT: 7.76). This represents a significant performance gain over the LLM-only method, which, despite a respectable 66.2\% success rate, suffered from lower and more inconsistent quality scores. By contrast, Direct prompting proved unsuitable for practical use, with a low 29.2\% success rate and collapsed scores across all quality dimensions.

The remainder of this paper is organized as follows. 
Section~\ref{sec:background} defines the experimental framework by specifying the type of pipelines under study, introducing the adopted component model, and outlining the dimensions considered for evaluation. 
Section~\ref{sec:pipeline_case_studies} describes the case studies addressed in the experiments.
In Section~\ref{sec:relatedWorks}, we review prior research on LLM-based code generation and workflow automation. 
Section~\ref{sec:Methodology} presents our Prompt2DAG methodology in detail, covering each stage from pipeline analysis to automated evaluation. 
Next, Section~\ref{sec:experiment} describes the experimental setup, and discusses our key findings and comparative analyses. 
Finally, Section~\ref{sec:conclusions} summarizes our contributions and outlines promising directions for future research in automated data pipeline development.
\color{black}

\section{Background}
\label{sec:background}

This paper investigates whether, and to what extent, LLMs can be used to generate executable pipelines from natural language descriptions. The objective is to develop a set of best practices and reusable templates to assist domain experts with limited technical expertise in automating their workflows. In this section, we present the context in which the solution was developed, along with the SemT service model assumed in the pipeline design. Finally, the approach to pipeline generation and validation is outlined.

\subsection{Problem definition}

To develop, demonstrate, and evaluate our methodology, we focus on data preparation pipelines as a representative and broadly applicable case study.
Data preparation is relevant in many domains, ranging from analytics to training of AI models.
In particular, we focus on tabular data enrichment, a critical phase in data preparation that aims to augment or extend existing data with additional information from external sources to enhance its value. 

A typical tabular data enrichment workflow begins with \textit{data cleaning}, which corrects errors and inconsistencies to ensure high-quality underlying data~\cite{bandiera2024using}, followed by \textit{data formatting}, which transforms data to comply with standards and facilitate subsequent processing~\cite{sajid2019predictive}.
The key task in data enrichment is \textit{data reconciliation}, which matches source data with data in a shared dataset, typically a knowledge graph, to enable the retrieval of data from another dataset and augment the source table with contextual information. This last task is known as \textit{data extension}, which involves adding new columns to the source tables to prepare the data for downstream tasks, such as analytics or training AI models.

\begin{figure}[h]
\centering
\includegraphics[width=0.95\columnwidth]{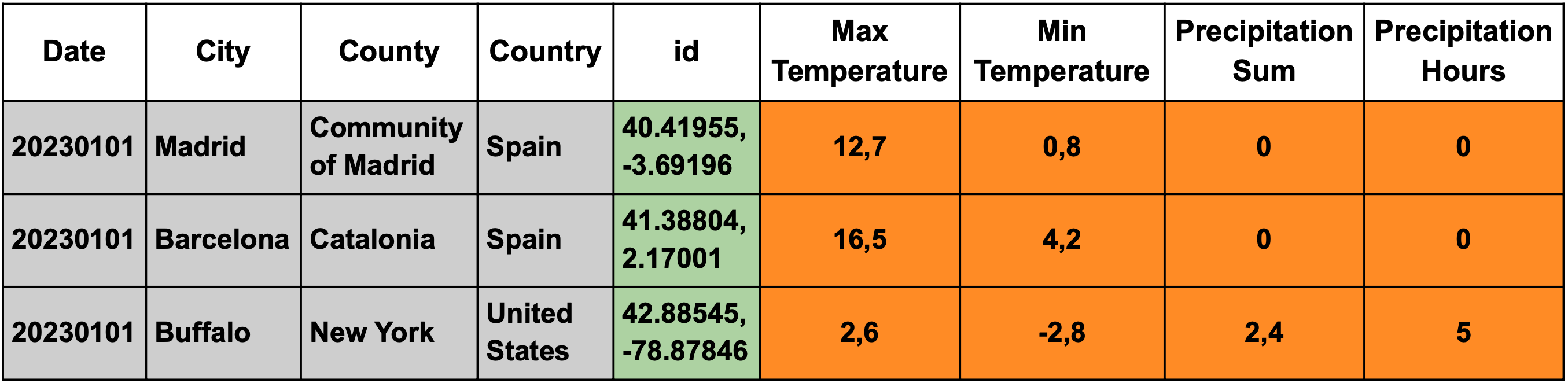}
\caption{Example of data enriched with geolocation and weather data}
\label{fig:enriched_data_example}
\end{figure}

Such workflows typically involve integrating heterogeneous data, applying complex transformations, and managing dependencies between diverse processing steps, making them an ideal testbed. Figure~\ref{fig:enriched_data_example} illustrates a practical application where tabular data is enriched with geospatial and meteorological information. The original table (gray) is augmented with geographical coordinates (green), which, combined with date information, enable the retrieval of weather data (orange). This process exemplifies common pipeline operations such as data reconciliation (e.g., city names), data type standardization (e.g., dates), and external API integration.

The objective is to automate these workflows, which are traditionally executed manually by data scientists who combine tools and services to prepare data for downstream tasks.
The proposed methodology enables the construction of pipelines, represented as Directed Acyclic Graphs (DAGs), from natural language descriptions by defining prompt templates that guide LLMs in generating structured enrichment pipelines.

A peculiar aspect of enrichment is leveraging external APIs and services, for example, geocoding and weather services, to inject contextual information that augments temporal, spatial, and environmental understanding~\cite{krasteva2023geospatial, Santos}. After such external enrichment, \textit{data integration} processes merge the augmented data with the source datasets while maintaining accuracy and utility~\cite{alotaibi2023cleaning}.

\subsection{SemT-Model}
\label{sec:semt}
In this work, we adopt the SemT-model~\cite{alidu2025semt}, based on the SemT architecture~\cite{35_RipamontiDePaoliPalmonari2022}, which provides a structured approach for constructing scalable and efficient multi-step data enrichment pipelines. The SemT model categorizes services and defines a backend architecture that facilitates the integration of diverse data sources and services, ensuring modularity, flexibility, and abstraction within enrichment workflows. Advantages are discussed in Table~\ref{tab:SemT-adv}.

\begin{table}[th]
\centering
\scriptsize
\begin{tabular}{>{\raggedright\arraybackslash}p{2.2cm}>{\raggedright\arraybackslash}p{10.5cm}}
\toprule
\textbf{Property} & \textbf{Description} \\
\midrule
\textbf{Flexibility} & Facilitates the integration of both new and existing services through APIs. This is particularly relevant because, unlike data integration, data enrichment typically relies on external, uncontrolled datasets accessible only via APIs. \\
\textbf{Standardization} & Promotes the use of common data models and protocols to ensure interoperability. Shared models simplify the definition of generic templates that can be applied across diverse scenarios. \\
\textbf{Abstraction} & Hides underlying service complexity to improve replaceability and simplify the development and maintenance of enrichment pipelines. Replaceability enables the evaluation of non-functional properties, such as accuracy and cost, allowing the selection of services that best meet technical and business requirements. \\
\textbf{Modularity} & Supports the reuse and composition of enrichment pipeline components. A shared service and data model facilitates the encapsulation of diverse logic and services behind standardized interfaces and behaviors, enabling automated management. \\
\textbf{Scalability} & Enables independent orchestration of components to support horizontal scaling. Pipelines can be decomposed into autonomous components, allowing for the parallel execution of time-critical tasks. \\
\bottomrule
\end{tabular}
\caption{SemT model advantages.}
\label{tab:SemT-adv}
\end{table}

\begin{figure}[th]
\centering
\includegraphics[width=0.55\textwidth]{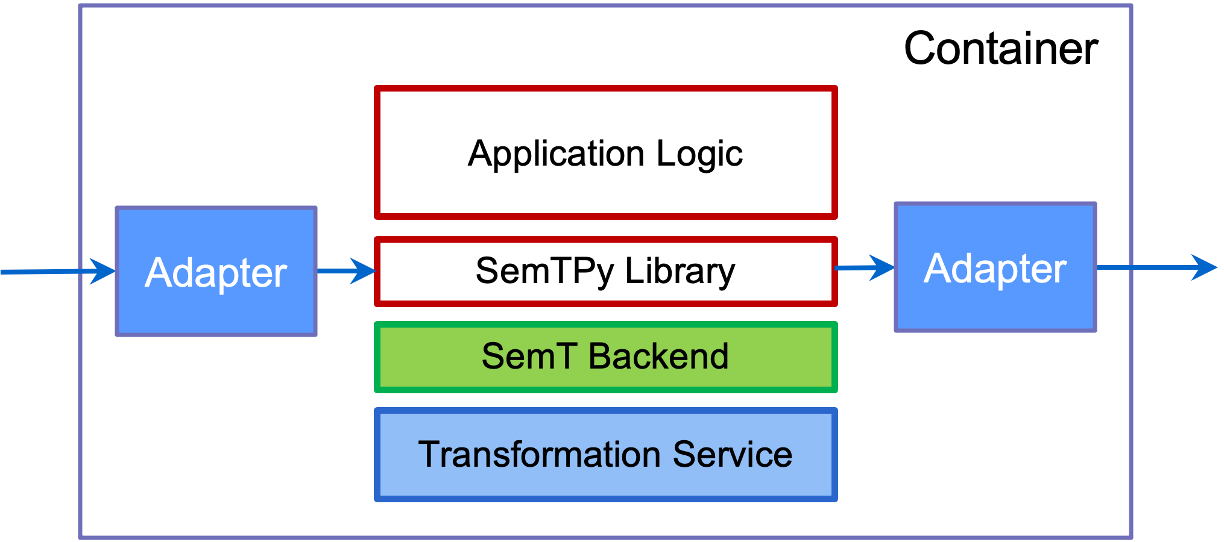}
\captionsetup{justification=centering}
\caption{An example of a container for data transformation.\\ SemT-Py library, SemT-backend and Adapters are standard components. }
\label{fig:dockerimage}
\end{figure}

The SemT model decouples client applications from specific service logic. The outcome is the definition of autonomous containers that can be used as building blocks in pipelines.
The core module in a container is a \texttt{SemT-backend} instance that acts as a standard advanced gateway, offering a unified Web API for seamless access to heterogeneous enrichment services. The SemtPy Python library provides the application logic with programmatic access to the backend and support inter-container communication via adapters (see an example in Figure~\ref{fig:dockerimage}). Adapters send and receive augmented tabular data complaint with the W3C API specification recommendation~\cite{w3cAPI} to enhance interoperability and composability via standards.

\subsection{Pipeline creation and validation}
We adopted a modular approach in defining the Prompt2DAG methodology, which offers the advantage of decomposing the complex task of pipeline generation into distinct, manageable stages. This approach aligns with Model-Driven Engineering (MDE) principles~\cite{burgueno2025automation} and follows software development best practices that emphasize separation of concerns~\cite{dixit2024systematic} and iterative refinement~\cite{liu2024large}. 

A modular approach not only improves the robustness and precision of the generated artifacts but also facilitates hierarchical prompt composition for the LLM-driven stages, allows for component-level optimization, enables multi-stage validation, and simplifies error localization. Furthermore, it inherently supports the human-in-the-loop paradigm ~\cite{natarajan2025human}, where users can inspect, validate, and potentially refine the intermediate outputs before proceeding to the next stage, offering greater control and trust in the automated process~\cite{tsamados2024human}.

Evaluating the quality of generated code, particularly for complex systems like data pipelines, requires multi-dimensional assessment. Traditional software quality metrics focus on aspects like correctness, maintainability, and performance \cite{lee2014software}. However, these may not fully capture the unique requirements of workflow orchestration code.

Recent work has proposed specialized frameworks for evaluating LLM-generated code, assessing dimensions such as functional correctness, adherence to best practices, and security \cite{kharma2025security}. However, these often focus on standalone programs or code snippets rather than integrated, multi-component systems, like data pipelines.

For workflow orchestration specifically, evaluation should consider code quality, structural integrity, and executability. Code quality refers to the readability, maintainability, adherence to coding standards, complexity, and security of the generated script. Structural integrity assesses the correctness of the DAG structure, including acyclicity, connectedness, task dependencies, and proper operator configurations. 
Executability measures the compatibility of the generated DAG with the target orchestrator (e.g., Airflow or ArgoFlow), ensuring that it can be loaded, parsed, and simulated correctly.

To address these dimensions, we adopt a Static Code Analysis Test (SAT) to evaluate code quality in terms of style compliance, complexity, and security; a DAG Structural and Configuration Analysis Test (DST) to examine the structural properties of the DAG, including graph acyclicity, task dependencies, and the correctness of operator configurations; and a Platform Conformance Test (PCT) to verify that the generated DAG can be loaded and parsed correctly, and that tasks can be simulated via dry-run execution (details in Section~\ref{sec:automatedEvaluation}).







\section{Related Work} \label{sec:relatedWorks}

This section examines the relevant literature focusing on the challenges raised by automated data pipeline generation and the evolution from template-based to LLM-driven approaches. 

\subsection{Data Pipeline Orchestration and Generation Challenges}

Data pipelines are structured sequences of data processing elements designed to extract, transform, and load data from various sources into target systems~\cite{tiwari2023data}. Modern organizations increasingly rely on these automated workflows to manage complex data processes across domains, including business intelligence, scientific computing, and machine learning operations~\cite{sresth2023optimizing, Nara12134}.

Several prominent workflow orchestration frameworks address these needs, each with distinct characteristics. Apache Airflow has emerged as a leading open-source solution, widely adopted for its flexibility, scalability, and Python-centric approach to defining workflows as Directed Acyclic Graphs (DAGs)~\cite{singh2019airflow, harenslak2021data}. Other notable frameworks include Luigi (Spotify's dependency-focused solution), Prefect (offering dynamic workflow capabilities), Dagster (emphasizing data-aware orchestration), and cloud-native solutions like AWS Step Functions and Azure Data Factory~\cite{luigi, prefect, dagster}.
Despite their power, these frameworks present significant barriers to entry, requiring specialized knowledge of programming, system integration, and workflow design patterns that often prevent domain experts from directly creating workflows without engineering support~\cite{Nara12134, Dureja2024}.


\subsection{Evolution from Template-Based to LLM-Driven Code Generation}

Traditional approaches to automated pipeline generation have relied heavily on template-based code generation (TBCG), a well-established technique that uses predefined templates with placeholders to produce code~\cite{syriani2018systematic}. In data engineering, template engines such as Jinja2 are widely used; for example, Apache Airflow supports Jinja2 templating within DAG definitions to dynamically incorporate values~\cite{astronomer2025templating, jinja2025}. TBCG offers predictability and control, as the generated code is derived directly from known templates, which simplifies auditing and debugging. However, the need to manually craft templates for each use case, combined with the limited logic supported by most template engines, represents a significant limitation~\cite{fastercapital_pipeline_generation}, making TBCG suitable only for regular and repetitive patterns~\cite{syriani2018systematic}.

The emergence of LLMs has introduced new possibilities for code generation, with models demonstrating remarkable capabilities in generating code from natural language prompts~\cite{srivatsa2024survey, wang2023review}. Unlike TBCG's fixed templates, LLMs are probabilistic and context-driven, synthesizing code by learning patterns from massive code and text corpora~\cite{wang2023review}. LLM-driven tools are increasingly applied to auto-generate pipeline code, with users prompting models for workflow descriptions and receiving complete scripts~\cite{astronomer_airflow_cohere_2025}. However, generating complex, multi-component systems like data pipelines presents unique challenges including maintaining consistency across components, ensuring proper integration and data flow, and adhering to framework-specific best practices~\cite{yasmin2025empirical}. Research indicates that LLMs, especially in direct end-to-end generation, can struggle with these aspects, often producing code that appears plausible in isolation but fails when integrated into larger systems~\cite{zamfirescu2023johnny, chen2024empirical}.

Existing LLM-based approaches typically employ monolithic generation strategies that struggle with the complexity and precision required for production-ready data pipelines. The lack of structured decomposition and intermediate validation leads to reliability issues and makes error diagnosis difficult.

\subsection{Advanced Prompting and Modular Design Principles}

Recent advances in prompt engineering demonstrate that task presentation significantly impacts LLM output quality, with structured prompting approaches showing particular promise for complex tasks~\cite{li2024acecoder}. Chain-of-Thought (CoT) prompting guides LLMs through intermediate reasoning steps to improve performance on complex reasoning tasks~\cite{wei2022chain}, while decomposed prompting breaks complex tasks into simpler subtasks, aligning with divide-and-conquer paradigms~\cite{khot2022decomposed}. Few-shot learning with exemplars provides LLMs with examples demonstrating desired reasoning processes and output formats~\cite{wang2020generalizing, madaan2022language}. However, the systematic application of these techniques to code generation for complex data pipelines, particularly in multi-stage frameworks like data enrichment pipelines, remains underexplored.

Software engineering principles for managing complexity provide additional guidance for LLM-based system generation. Separation of concerns advocates dividing complex systems into distinct sections with minimal overlap to reduce cognitive load~\cite{dixit2024systematic}, while progressive refinement involves developing systems through successive levels of detail, starting with high-level abstractions~\cite{liu2024large}. Intermediate representations serve as bridges between abstraction levels, capturing essential information while hiding unnecessary details~\cite{chakroborti2022gain}. These principles, while well-established in traditional software development, have not been systematically applied to LLM-based workflow generation.

Current literature lacks comprehensive methodologies that combine advanced prompting techniques with software engineering principles specifically for data pipeline generation. Most existing approaches focus either on general code generation or simple workflow creation, without addressing the specific challenges of data enrichment pipelines that require external service integration, data transformation orchestration, and business domain expertise.


\section{Case Studies}
\label{sec:pipeline_case_studies}

To evaluate the Prompt2DAG methodology within the data enrichment pipeline domain, we provide five representative scenarios that demonstrate how different business contexts require distinct enrichment strategies and external service integrations. While these pipelines share a common sequential or parallel architecture typical of data enrichment workflows, they vary significantly in their business objectives, data processing requirements, and external service dependencies.

Our proposal specifically targets data enrichment pipelines, that is workflows that augment existing datasets with additional information from external sources. This domain encompasses common enterprise scenarios where organizations need to enhance their data assets through entity reconciliation, sentiment analysis, geocoding or other value-adding transformations. The selected case studies represent distinct business applications, each requiring different enrichment strategies and service integrations.

\begin{sidewaystable}[htbp]
\centering
\caption{Comprehensive comparison of all five data enrichment pipeline case studies, showing domain, enrichment goal, processing and parallelism pattern, external service integration, and complexity}
\label{tab:pipeline_case_studies_comprehensive}
\scriptsize
\begin{tabular}{p{0.16\textwidth} p{0.14\textwidth} p{0.18\textwidth} p{0.16\textwidth} p{0.14\textwidth} p{0.14\textwidth}}
\toprule
\bf{Characteristic} 
   & \bf{DM Sequential} 
   & \bf{DM Pipeline-level Parallelism} 
   & \bf{DM Task-level Parallelism} 
   & \bf{Procurement Supplier Validation} 
   & \bf{Multilingual Product Review}\\
\midrule
\bf{Business Domain} 
   & Marketing Analytics 
   & Marketing Analytics 
   & Marketing Analytics 
   & Procurement/Supply Chain 
   & E-commerce/Text Analytics \\
\addlinespace
\bf{Enrichment Objective} 
   & Geographic and meteorological data augmentation for customer targeting 
   & Same, but at high throughput via parallel chunks 
   & Same, with reconciliation accelerated by internal API concurrency 
   & Supplier info validation/standardization 
   & Multilingual review content/sentiment enrichment \\
\addlinespace
\bf{Primary Services} 
   & HERE Geocoding, OpenMeteo 
   & HERE Geocoding, OpenMeteo (concurrent per parallel branch) 
   & HERE Geocoding (with high-concurrency), OpenMeteo 
   & Wikidata/Wikibase reconciliation 
   & Language detection, LLM inference\\
\addlinespace
\bf{Service Integration Pattern}
    & Strictly sequential: Load $\rightarrow$ Geocode $\rightarrow$ Weather $\rightarrow$ ColumnExt $\rightarrow$ Save
    & "Scatter-gather": Split $\rightarrow$ (5 steps $\times$ N in parallel) $\rightarrow$ Sync $\rightarrow$ Merge
    & Sequential, but heavy tasks (e.g. geocode) are implemented as multi-threaded/multi-request
    & Load $\rightarrow$ Entity Reconciliation $\rightarrow$ Save
    & Load $\rightarrow$ LangDetect $\rightarrow$ Sentiment $\rightarrow$ Feature Extraction $\rightarrow$ Save \\
\addlinespace
\bf{Data Processing Steps}
    & 5: Load, Geocode, Weather, ColExt, Save
    & 7(+): Split, [parallel: Load, Geocode, Weather, ColExt, Save] x N, Sync, Merge
    & 5: Load, Geocode (internally parallel), Weather, ColExt, Save
    & 3: Load, Reconcile, Save
    & 5: Load, LangDetect, Sentiment, Feature, Save \\
\addlinespace
\bf{Parallelization Level}
    & None (strict order)
    & Pipeline-level (across independent data chunks)
    & Task-level (in reconciliation/geocoding task only)
    & None
    & None \\
\addlinespace
\bf{External Dependencies}
    & HERE API, OpenMeteo API
    & HERE API, OpenMeteo API (multiple branches)
    & HERE API (high QPS), OpenMeteo API
    & Wikidata SPARQL endpoint
    & LangDetect service, LLM endpoints (HuggingFace, etc.) \\
\addlinespace
\bf{Enrichment Complexity}
    & Coordinate-based weather lookup
    & Same complexity, distributed over parallel tasks
    & Complexity hidden by concurrent API invocations
    & Knowledge base entity linking
    & Multi-step NLP pipeline \\
\addlinespace
\bf{Use-case Fit}
    & Small/medium, legacy, traceable processing
    & High-volume campaigns, time-sensitive bulk jobs
    & Geocoding performance-intensive, API rate-limited scenarios
    & Supplier data cleansing
    & Global review analytics \\
\bottomrule
\end{tabular}
\end{sidewaystable}

Table~\ref{tab:pipeline_case_studies_comprehensive} presents a comprehensive comparison of five data enrichment pipeline scenarios, including the baseline sequential, pipeline-level parallel, and task-level parallel Digital Marketing cases, alongside the Procurement and Multilingual Review cases. Each scenario highlights distinct business domains, processing patterns, enrichment complexity, external integrations, and parallelization strategies.
The case studies demonstrate three primary approaches to \textit{external service integration} typical in modern data enrichment pipelines:

\begin{itemize}[label={},leftmargin=0pt,itemindent=1.5em]
    \item \textbf{Geographic Services Integration (Digital Marketing):} Utilizes HERE Geocoding API for location standardization, followed by OpenMeteo API for weather data retrieval—representing coordinate-based enrichment patterns common in location-aware business applications.
    
    \item \textbf{Knowledge Base Reconciliation (Procurement):} Employs Wikidata's SPARQL endpoint for entity linking and supplier validation, exemplifying structured knowledge integration in supply chain and procurement scenarios.
    
    \item \textbf{NLP Service Orchestration (Multilingual Product Review):} Leverages language detection services, together with LLM-based sentiment analysis and feature extraction, reflecting multi-step NLP processing chains in content analytics.
\end{itemize}

Each scenario addresses distinct organizational needs within the data enrichment domain. The Digital Marketing pipeline supports location-based customer segmentation by augmenting customer data with geographic and weather information. The Procurement pipeline enables supplier data quality improvement through entity standardization and validation. The Multilingual Review pipeline facilitates international e-commerce insights by processing customer feedback across multiple languages.

Within the Digital Marketing domain, we further explore alternative architectural implementations of the same enrichment logic, distinguished by their parallelization strategy:

\begin{itemize}[label={},leftmargin=0pt,itemindent=1.5em]
    \item \textbf{Pipeline-level Parallelism (Digital Marketing, Parallel Branches):} This advanced variant partitions input data into multiple chunks, each processed in parallel by an independent pipeline branch (Docker container) implementing the same enrichment sequence. This "scatter-gather" pattern synchronizes and merges results after all branches complete, dramatically boosting throughput for large-scale datasets.

    \item \textbf{Task-level Parallelism (Digital Marketing, Internal Service Concurrency):} In this case, the overall pipeline remains sequential, but the most computationally intensive step—geocoding—is internally parallelized within a single container. By issuing concurrent API requests to the HERE Geocoding service, this approach accelerates enrichment without restructuring the pipeline itself.
\end{itemize}


The business objectives and external integrations are the same across the three Digital Marketing scenarios, their architectural choices for parallelization—either at the pipeline level or within individual tasks—demonstrate the adaptability of data enrichment workflows to different scalability and performance requirements.


These scenarios provide a focused evaluation framework for assessing how well Prompt2DAG can interpret business-specific natural language descriptions and generate appropriate data enrichment workflows, while maintaining clear boundaries within the data enrichment pipeline domain.

All pipeline scenarios are implemented using containerized components that encapsulate specific enrichment operations, as illustrated in Figure~\ref{fig:dockerimage}. Each component is designed to perform a well-defined transformation or enrichment task. These containers are developed according to the \texttt{SemT-model}, introduced in Section~\ref{sec:semt}, to provide standardized and reusable building blocks for data processing.

All case studies follow consistent architectural patterns that reflect diverse data enrichment workflows across multiple business domains. Each pipeline implements containerized services that transform data through standardized operations: data ingestion, external service integration, enrichment processing, and persistence. 
These scenarios collectively establishes the user stories for evaluating Prompt2DAG's capacity to generate diverse enrichment architectures from natural language specifications at the level of both orchestration and enrichment logic. They systematically cover sequential processing, pipeline-level parallelism (scatter-gather), task-level concurrency, entity reconciliation workflows, and multi-step NLP pipelines, thereby testing both orchestration-level optimization and microservice-level performance strategies across realistic business use cases.

Table~\ref{tab:data_enrichment_components} presents the core component types leveraged across the data enrichment scenarios.






\begin{table}[th]
\centering
\scriptsize
\begin{tabular}{>{\raggedright\arraybackslash}p{3cm}>{\raggedright\arraybackslash}p{10cm}}
\toprule
\textbf{Component Type} & \textbf{Description} \\
\midrule
\textbf{Load and Modify Data} & Initial component responsible for ingesting raw data from CSV files, performing modification operations, such as data format normalization (e.g., converting dates to ISO 8601), and transformation to structured JSON format suitable for subsequent pipeline stages. \\
\textbf{Data Reconciliation} & Critical component for data quality that standardizes and disambiguates entities within the dataset. Receives JSON data and outputs enriched JSON with reconciled information, implementing either entity linking through knowledge bases (Wikidata API for supplier validation) or geospatial reconciliation through location services (HERE Geocoding API for coordinate resolution). \\
\textbf{Data Extension} & Components that augment datasets with new information derived from existing or reconciled data. This includes meteorological data extension (OpenMeteo API for weather information), NLP-based enrichment (language detection and sentiment analysis), and custom column extensions based on business logic requirements. \\
\textbf{Save Data} & Final component responsible for persisting the fully enriched dataset, typically converting from JSON back to CSV format or other specified output formats in designated data directories. \\
\textbf{Split Dataset} & Orchestration component that partitions the raw input file into multiple independently processable segments, enabling parallel execution of downstream tasks. \\
\textbf{Synchronize and Merge} & Orchestration component that ensures all parallel branches have completed processing and consolidates their outputs into a unified dataset for subsequent storage or analysis. \\
\bottomrule
\end{tabular}
\caption{Core component types for data enrichment scenarios}
\label{tab:data_enrichment_components}
\end{table}

The containerized architecture operates under several key assumptions that define a consistent execution model that provides the foundation for automated pipeline generation:

\begin{itemize}[label={},leftmargin=0pt,itemindent=1.5em]
    \item \textbf{Orchestration Framework:} Apache Airflow serves as the workflow orchestrator, with pipeline tasks executed using DockerOperator for containerized execution and sequential task dependencies managed through Airflow's DAG structure.

    \item \textbf{Container Infrastructure:} All pipeline components execute within Docker containers that share a persistent data volume for file-based data exchange and utilize a managed Docker network for service communication.

    \item \textbf{Service Architecture:} Pipeline components follow a client-server model where containerized Python scripts act as clients connecting to pre-deployed backend services (SemT-backend instances) that provide the actual enrichment functionality through standardized APIs.

    \item \textbf{Pre-built Components:} The methodology assumes availability of pre-built Docker images containing both the Python client scripts for each pipeline step and the necessary backend services, with service lifecycle management handled externally through Docker Compose. 

    \item \textbf{Configuration Management:} Task parameters and service configurations are passed through environment variables and command-line arguments, with external service credentials and API keys managed through Airflow's configuration system.
\end{itemize}


\section{Prompt2DAG}\label{sec:Methodology}

\begin{figure}[!ht]
    \centering
    \includegraphics[width=0.95\textwidth]{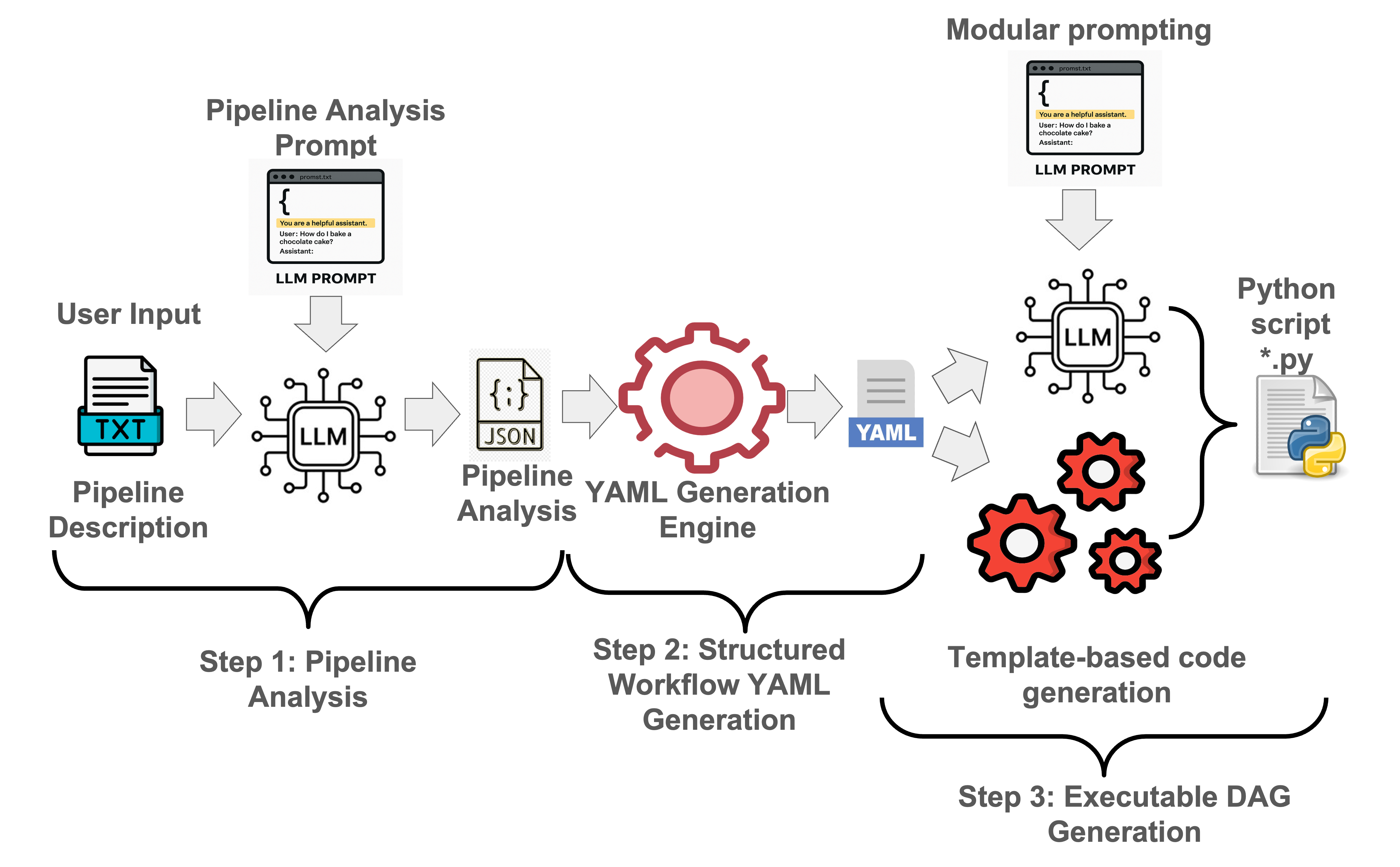}
    \caption{Overview of the Prompt2DAG Methodology}
    \label{fig:Prompt2dag_arch}
\end{figure}

This section details the \textbf{Prompt2DAG methodology}. The core principle of Prompt2DAG as illustrated in Figure~\ref{fig:Prompt2dag_arch} is a phased transformation:
\begin{enumerate}
    \item \textbf{Pipeline Analysis.} A natural language description of the desired pipeline is analyzed using a chain of prompts to extract a comprehensive, structured JSON representation detailing components, data flow, parameters, and integrations.\label{meth:pipeline_analysis}
    \item \textbf{Structured Workflow Generation.} The structured JSON artifact is deterministically transformed into a standardized, platform-neutral workflow specification, enhancing readability and serving as a stable intermediate representation.\label{meth:structured_workflow_generation}
    \item \textbf{Executable DAG Generation.} The workflow specification is then used to automatically generate executable DAG code for a target orchestration platform (e.g., Apache Airflow), using either \begin{enumerate*}[label=(\alph*), ref=\theenumi.\alph*]
        \item LLM-driven modular synthesis or\label{meth:executable_dag_generation_a}
        \item deterministic template-based expansion.\label{meth:executable_dag_generation_b}
    \end{enumerate*}  
    \item \textbf{Automated Evaluation.} The generated DAG code undergoes rigorous automated evaluation, assessing code quality, structural integrity, and dry-run executability to ensure reliability and correctness.\label{meth:automated_evaluation}
\end{enumerate}

By decomposing the problem into these distinct, well-defined stages, Prompt2DAG emphasizes transparency, reproducibility, and modularity. Each step involves precisely defined inputs, processes (including verbatim LLM prompts where applicable), and output schemas. 
The subsequent sections will elaborate on each of these steps in detail.

\subsection{Step 1: Pipeline Analysis}\label{sec:prompt2dag-analysis}

The Pipeline Analysis phase transforms pipeline descriptions into structured, machine-interpretable representations. This transformation presents a fundamental challenge: descriptions often interweave technical specifications with business logic, combine sequential and parallel execution patterns, and reference external systems using domain-specific terminology. 

Our approach employs task decomposition and prompt chaining techniques that break the analysis into focused sub-tasks, each designed to extract specific aspects of the pipeline architecture. The analysis begins with a plain-text description that captures objectives, processing steps, data flow patterns, configuration parameters, and external system interactions. 

Each analytical sub-task follows a consistent pattern: a carefully crafted prompt guides the LLM to examine the description from a particular perspective, extract relevant information, and format it according to a predefined schema. The structured output from one task becomes input for subsequent tasks, creating a chain of analytical steps that progressively builds a complete pipeline model. This approach improves accuracy by focusing the LLM on narrower objectives, enhances reproducibility through discrete prompt-response pairs, and provides transparency by making each analytical decision traceable.

The Pipeline Analysis unfolds through five stages, each building upon previous results to construct a comprehensive structural representation. 
The process begins with \textbf{environment inference} (see  \ref{subsec:env_prompt} for full prompt), where the system analyzes contextual cues within the description to determine the intended execution platform. 

The following step is the \textbf{component identification}, which extracts the fundamental building blocks of the pipeline (see  \ref{subsec:comp_type_prompt} for full prompt). 
The classification employs a standardized taxonomy that encompasses data-centric operations (DataLoaders for ingestion, Transformers for manipulation), coordination tasks (Reconciliators for standardization, Enrichers for augmentation), output operations (Exporters, QualityChecks), and control flow elements (Splitters, Mergers, Orchestrators). 
Details on the component types and classification are provided in \ref{appendix:component_types}.

The \textbf{flow structure analysis} stage assembles these identified components into a complete execution graph. 
The system determines which components serve as pipeline entry points, traces the connections between sequential tasks, and identifies sections where multiple instances of a component sequence execute in parallel (see  \ref{subsec:flow_prompt} for full prompt). 

\textbf{Parameter extraction} follows, systematically identifying all configurable aspects mentioned in the pipeline description. This stage distinguishes between global parameters that affect the entire pipeline (such as data directories or API tokens), component-specific parameters that configure individual processing steps, and environment variables that provide runtime configuration. For each parameter, the system extracts its data type, default value if specified, requirement status, and any mentioned constraints or validation rules (see  \ref{subsec:para_prompt} for full prompt). 

The \textbf{integration analysis} stage identifies all points where the pipeline interacts with external systems. This includes APIs, databases, file systems, message queues, and cloud services. For each integration point, the system captures connection details, authentication requirements, the components that interact with it, and the direction of data flow. Additionally, this stage identifies the pipeline's primary data sources and ultimate data sinks (see  \ref{subsec:int_prompt} for full prompt). 



The Pipeline Analysis produces a human-readable report and a comprehensive JSON artifact. The report covers architecture, components, parameters, and integrations, serving as documentation and validation for pipeline authors (prompt templates in  \ref{appendix:prompt_templates}). The JSON artifact follows a versioned schema (currently v1.3) and includes analysis metadata, pipeline summary, component definitions, flow structure with parallel patterns, scoped parameters, and integration specifications. It serves as primary input for subsequent workflow generation (schema and example in \ref{appendix:Step_1_JSON}).

Beyond these primary outputs, the system generates auxiliary artifacts for debugging, validation, and reproducibility. 



\subsection{Step 2: Structured Workflow Generation}
\label{sec:prompt2dag-yamlgeneration}

The second phase of Prompt2DAG performs a deterministic model-to-model (M2M) transformation~\cite{favre2004foundations} that converts the analysis-centric JSON artifact from Step \ref{meth:pipeline_analysis} into a standardized, platform-neutral workflow specification. This intermediate step addresses fundamental limitations in direct LLM-to-code generation while establishing a clean abstraction layer between pipeline analysis and platform-specific implementation.

The transformation from the analysis JSON artifact to YAML workflow specification format serves multiple strategic purposes: normalizing complex nested structures into a linear, human-readable format with explicit control flow constructs, performing semantic reorganization from analysis-centric to execution-centric organization, and maintaining semantic richness while optimizing for downstream transformation. 

The use of YAML is motivated by both empirical and architectural concerns. 
YAML’s indentation-based syntax and reduced reliance on brackets have helped minimize inconsistencies produced by LLMs, while also supporting readability and facilitating review.
Moreover, YAML is widely adopted for workflow and infrastructure configuration (e.g., Kubernetes, GitHub Actions, Ansible), aligning with patterns familiar to practitioners in data engineering and MLOps. 
Finally, YAML’s more concise syntax may reduce token usage compared to equivalent JSON representations. 

The generated workflow specification follows the schema detailed in \ref {appendix:YAML_Schema}.


\subsection{Step 3: Executable DAG Generation}
\label{sec:prompt2dag-daggeneration}

The final transformation phase converts the platform-neutral workflow specification into concrete, executable assets (e.g., Python scripts) for target orchestration platforms, like Apache Airflow. 
%
Prompt2DAG offers two complementary pathways: 

\textbf{Template-Based Generation}, which parses the workflow specification file into context dictionaries, applies custom template filters for platform-specific formatting, and renders complete scripts with metadata headers, imports, operator definitions, and dependency declarations. 

\textbf{LLM-powered Generation}, which employs prompt chaining with few-shot learning, guiding language models through separate phases: DAG structure generation (imports, arguments, context manager), individual operator instantiation for each workflow task, and dependency translation into platform-specific syntax. Each phase uses carefully crafted prompts with platform documentation examples for In-Context-Learning \cite{dong2022survey}.

Both pathways incorporate comprehensive validation for syntactic correctness, extensive metadata headers for traceability (source YAML, timestamps, generation method, warnings), and preserve all parameter specifications from global pipeline settings to component-specific configurations using appropriate platform mechanisms (see  \ref{appendix:DAG_Generation} for full details on prompts and sample outputs).

\subsection{Step 4: Automated Evaluation}\label{sec:automatedEvaluation}

The final step uses a structured platform-specific (in our case, Apache Airflow) evaluation framework combining static analysis, graph-theoretic checks, and platform-specific simulations. This framework quantitatively assesses generated DAG code for correctness, maintainability, structural integrity, and executability.

Following best-practice, we decouple basic loadability (a hard binary gate) from the richer notion of runtime executability. For each generated DAG, we define a binary variable

\[
\texttt{executability\_score} \in \{0, 1\}
\]

which is 1 if the DAG is loadable by Airflow's DagBag, 0 otherwise. All downstream quality scores (SAT, DST, PCT) are computed for all samples, with non-loadable DAGs assigned a score of zero. This ensures that unreliable methods are penalized in aggregate statistics and means.

The evaluation proceeds with the calculation of the following three metrics:
\begin{itemize}[label={},leftmargin=0pt,itemindent=1.5em]
\item \textbf{Static Code Analysis Test (SAT):} 
This phase assesses Python code quality based on adherence to coding standards (PEP 8), style consistency, syntactic correctness, security vulnerabilities, and complexity. 
Tools include Pylint, Flake8, Bandit, and Radon. A weighted penalty model is used:
\begin{equation*}
\text{SAT} = \text{Normalize} \left( \sum_{i=1}^{4} w_i \cdot s_i \right)
\end{equation*}
where $w_i$ is the weight for each dimension (Security, Pylint, Flake8, Complexity) and $s_i$ is the corresponding penalty-adjusted score.

\item \textbf{DAG Structural and Configuration Analysis Test (DST):} 
This component checks task and dependency structure, acyclicity~\cite{shi2023performance}, connectedness~\cite{kail2015achieving}, and depth/breadth~\cite{van2002alternative}, alongside operator configuration~\cite{straesser2023systematic}. The score is computed as:
\begin{align*}
S_{\text{struct}} &= 100 - \sum p_i \cdot v_i \\
S_{\text{config}} &= \text{Average}_{\text{operators}} \left( 100 - \sum q_j \right) \\
\text{DST} &= \frac{w_{\text{struct}} \cdot S_{\text{struct}} + w_{\text{config}} \cdot S_{\text{config}}}{10}
\end{align*}
where $p_i$ is the penalty for violation type $i$, $v_i$ is the number of violations of type $i$, $q_j$ is the penalty for configuration issue $j$, and $w_{\text{struct}} = 0.7$, $w_{\text{config}} = 0.3$.

\item \textbf{Platform Conformance Test (PCT):} 
This metric evaluates operational readiness and runnability on Apache Airflow:

\begin{enumerate}
  \item \textbf{Loadability:} The DAG must import successfully in Airflow’s \texttt{DagBag}.
  \item \textbf{Structural Gate (lenient):} The DAG must be acyclic and pass a
        connectivity check.\footnote{Cycles or zero tasks constitute a critical failure; isolated tasks merely generate warnings.}
  \item \textbf{Dry-Run Simulation:} Every task is executed with
        \texttt{airflow tasks test --dry-run}.  
\end{enumerate}

The score is calculated in two layers:

\begin{description}
  \item[Binary gate] 
  \[
  L = \mathbb{1}(\text{DAG is loadable via \texttt{DagBag}})
  \]
  If $L=0$ the DAG is unusable and all primary metrics (SAT, DST, PCT) are set to~0.

  \item[Weighted executability score]  (applied only when $L=1$)

  The evaluator assigns fixed base points for passing critical checks and a
  variable component for dry-run success:

  \[
     E = \min\!\Bigl( 10,\;
       3  \;+\;        
       2  \;+\;        
       B  \;+\;        
       4D \Bigr)       
  \]

  where
  \begin{itemize}
     \item $D = \texttt{successful\_dryrun\_ratio} =
            \dfrac{1}{N}\sum_{i=1}^{N}
            \mathbb{1}\bigl(\text{task}_i\text{ dry-run succeeds}\bigr)$,
           with $N$ the total number of tasks.  A task that fails
           \emph{only} because the local Docker daemon is unavailable is
           counted as a success (warning, not failure).
     \item $B$ is a small bonus reflecting overall structural quality,  
           \[
           B =
             \begin{cases}
               1.0 & \text{if } S_{\text{struct}} \ge 70\\[2pt]
               0.5 & \text{if } 50 \le S_{\text{struct}} < 70\\[2pt]
               0   & \text{otherwise,}
             \end{cases}
           \]
           where $S_{\text{struct}}\!(\!0\text{–}100)$ is the DST structure
           sub-score.\footnote{%
The constants 3 (loadability) and 2 (structure gate) partition the 0–10
scale so that \emph{passing the two critical gates but executing no tasks}
yields a mid-score of 5.  This reserves the upper 5 points for actual
task-level executability, while still discriminating pipelines that cannot
load or that violate essential structural rules.  The same weighting is
hard–coded in the evaluation script to keep paper and implementation
consistent.}
  \end{itemize}
  The constants (3 + 2) ensure that a DAG which loads and passes the structure
  gate but executes no tasks receives a baseline score of 5, while the
  dry-run band contributes up to four additional points.  The
  $\min(10,\cdot)$ cap keeps $E$ on the 0–10 scale.
\end{description}

\noindent
The final Platform Conformance Test score is 

\[
  \text{PCT} = L \times E .
\]

\end{itemize}

Table~\ref{tab:violations_penalties} summarizes the key parameters used across the three evaluation components (SAT, DST, and PCT), including default weights, penalties, and violations.

\begin{table}[H]
\centering
\scriptsize
\begin{tabular}{@{}lllll@{}}
\toprule
\textbf{Component} & \textbf{Violation Type} & \textbf{Penalty\,$\Delta$} & \textbf{Example} & \textbf{Rationale} \\
\midrule
\multirow{4}{*}{SAT} 
 & High-severity security issue     & $0.25$ & Hard-coded credentials      & Critical vulnerability \\
 & Medium Pylint error              & $0.05$ & Unused import               & Readability / maintainability \\
 & Minor PEP-8 style violation      & $0.02$ & Line length $>$ 79          & Low–impact cosmetic issue \\
 & High cyclomatic complexity       & $0.10$ & Function CC $>$ 10          & Maintainability risk         \\
\midrule
\multirow{6}{*}{DST} 
 & Cycle in DAG                     & $0.50$ & A → B → A                  & Breaks acyclicity, execution fails \\
 & Disconnected component           & $0.30$ & Isolated task sub-graph     & Part of DAG never triggered  \\
 & Isolated task                    & $0.10$ & Task with no deps           & Mild structural smell        \\
 & Missing required parameter       & $0.15$ & DockerOperator w/o image    & Task cannot run              \\
 & Invalid dependency               & $0.20$ & Edge to non-existent task   & Scheduling error             \\
 & Empty task group                 & $0.05$ & TaskGroup with 0 tasks      & Cosmetic configuration issue \\
\midrule
\multirow{4}{*}{PCT} 
 & DagBag loading failure           & $1.00$ & Syntax or import error      & DAG unusable; hard stop      \\
 & Airflow parsing warning          & $0.05$ & Deprecated param            & Non-blocking, future risk    \\
 & Task dry-run error               & $0.20$ & Invalid operator config     & Runtime failure for a task   \\
 & Platform validation error        & $0.50$ & Unsupported operator type   & DAG cannot execute properly  \\
\bottomrule
\end{tabular}
\caption{Violation catalogue and associated score deductions used in SAT, DST, and PCT. Penalties are expressed as absolute point deductions on the 0–10 scale before normalisation.}
\label{tab:violations_penalties}
\end{table}

In all aggregate reporting, non-loadable DAGs are assigned SAT, DST, and PCT scores of zero. Means and distributions in results thus reflect both method reliability and quality, ensuring honest assessment and direct comparability between competing methods. For example, a pipeline generation method that often produces invalid DAGs will have sharply reduced average SAT and DST scores, even if its successful outputs are high-quality. This approach makes comparison between generation methods both fair and transparent.

\subsection{Secondary Evaluation Metrics: LLM-Based Fidelity Assessment and Token Usage}
\label{sec:llm_quality_and_token}

Beyond structural, static, and executability (SAT, DST, PCT), we introduce secondary evaluation metrics that complement primary pipeline correctness and quality analyses:

\begin{itemize}
  \item \textbf{LLM-Based Step 1 Output Assessment:} 
  To objectively assess the \emph{semantic fidelity} and \emph{faithfulness} of the Pipeline Analysis output (Step 1), we adopt a Large Language Model (LLM)-as-a-judge paradigm. For each case, an external LLM (DeepSeek-chat) receives the pipeline description and downstream Step 1\&2 artifacts, plus a formal issue catalog and explicit system prompt. It is tasked to identify:
  \begin{itemize}
    \item \textbf{MISSING} information (present in description, absent in analysis)
    \item \textbf{HALLUCINATED} elements (present in outputs, absent in description)
    \item \textbf{INCONSISTENCIES} (mismatches in values, structure, or semantics)
    \item \textbf{CORRECT} (faithful, traceable representations)
  \end{itemize}

  The model must classify all major elements (components, parameters, integrations, workflow dependencies), reference a controlled vocabulary of issue types/codes (see Appendix~\ref{appendix:llm_assessor_issue_catalog}), and produce a structured JSON assessment with counts, severity, and evidence.

  A sample of the specific evaluation prompt and schema, as well as the full issue catalog and result template, is included in Appendix~\ref{appendix:llm_assessor_prompt}.

  Our analysis reports the following summary metrics per pipeline:
  \begin{itemize}
    \item Total number of critical fidelity issues (missing/hallucinated/inconsistent)
    \item Number of correctly identified items
    \item Severity histogram (HIGH/MEDIUM/LOW)
    \item Category breakdown (components, parameters, integrations, workflow logic)
  \end{itemize}

  This automatic LLM-based auditing creates a direct, reproducible, and scalable mechanism for semantic model validation before any code is generated.

  \item \textbf{LLM Token Usage Analysis.} 
  We separately track and report token usage (prompt, input, and output) for all LLM interactions in Prompt2DAG. This metric, computed using \texttt{tiktoken} and per-model encoders, provides an empirical measure of computational efficiency and operational cost across generative methods. This is essential both as a cost metric and as a measure of prompt ``weight'' which can impact generation reliability.

  Means and variances in total prompt/completion tokens are reported for each pipeline and method.
\end{itemize}

\section{Experimental Results}

\label{sec:experiment}

This section presents the experimental evaluation of the proposed methodology, examining both its technical performance and associated computational costs.
To ensure systematic analysis, we structure our evaluation around three complementary research questions that collectively assess the viability of automated DAG generation methodologies:

\textbf{RQ1: Performance Analysis} -- How do the four generation approaches (Direct, LLM-only, Hybrid, and Templated) compare in terms of reliability, structural quality, and executability across different pipeline complexities and LLM architectures? This establishes the fundamental performance baseline and identifies optimal generation strategies.

\textbf{RQ2: Pipeline Analysis Quality Assessment} -- How does the quality of initial prompt comprehension and pipeline analysis impact downstream generation success across different methodologies? This examines whether semantic fidelity in early stages serves as a reliable predictor of final DAG quality.

\textbf{RQ3: Computational Cost Analysis} -- What are the computational costs and cost-effectiveness trade-offs associated with each generation approach when accounting for both token consumption and success rates? This evaluates the economic viability of different methodologies for production deployment.

To address these research questions systematically, we developed a comprehensive experimental framework that ensures robust evaluation across multiple dimensions. 

\textbf{Target Pipelines.} All experiments utilize representative data enrichment pipeline scenarios discussed in Section~\ref{sec:pipeline_case_studies}. 

\textbf{Candidate LLMs.} Thirteen state-of-the-art LLMs were evaluated across diverse architectural paradigms and parameter scales. All models were configured with temperature=0.0 to ensure deterministic outputs. Table~\ref{tab:consolidated_benchmarks_expanded_condensed} consolidates their performance on established industry benchmarks, including HumanEval (HE) \cite{chen2021evaluating}, MBPP\cite{austin2021program}, MMLU~\cite{hendrycks2020measuring}, GSM8K~\cite{cobbe2021training}, Big Bench Hard (BBH), and GPQA. These benchmarks assess fundamental capabilities in reasoning, knowledge comprehension, mathematical problem-solving, and code generation that are directly relevant to DAG construction tasks.
The selected models span various architectural approaches: traditional dense transformers (Claude 3.5 Sonnet, GPT-4o Mini), mixture-of-experts architectures (DeepSeek-V3), and specialized models optimized for specific tasks (Phi-4 for reasoning, Mistral variants for multilingual capabilities). This diverse selection enables comprehensive analysis of how different model architectures and training methodologies impact performance on structured reasoning tasks such as DAG generation.

\begin{table}[ht]
\centering
\resizebox{\textwidth}{!}{%
\begin{tabular}{@{}lcccccccc@{}}
\toprule
Model & \#Params & Context & HE (p@1) & MBPP (p@1) & GSM8K & BBH & MMLU & GPQA \\
\midrule
DeepSeek-AI & 236B & 128K & 73.8\% & 75.4\% & 84.1\% & 82.3\% & 84.5\% & 55.2\% \\
Claude 3.5 Sonnet & -- & 200K & 92.0\% & -- & 92.0\% & 93.1\% & 88.7\% & 59.1\% \\
Meta Llama 3.3 70B (I) & 70B & 128K & 88.4\% & 87.6\% & 89.7\% & 85.4\% & 86.0\% & 50.5\% \\
Qwen2.5-72B (I) & 72.7B & 128K & 86.0\% & 80.2\% & 89.5\% & 82.4\% & 84.2\% & 37.9\% \\
DeepSeek-V3 & 671B & 128K & 65.2\% & 75.4\% & 89.3\% & 87.5\% & 88.5\% & 59.1\% \\
GPT-4o Mini & -- & 128K & 87.2\% & 81.3\% & 87.0\% & 85.2\% & 82.0\% & 53.6\% \\
Gemini 1.5 Flash & -- & 1M & 74.4\% & 71.9\% & 86.5\% & 78.9\% & 78.9\% & 48.9\% \\
Microsoft Phi-4 & 14B & 16K & 82.6\% & 76.8\% & 91.1\% & 68.1\% & 83.4\% & 52.8\% \\
Llama-4 Scout & -- & 10M & 69.5\% & 65.3\% & 82.7\% & 75.2\% & 78.1\% & 45.3\% \\
Mistral Large & 123B & 128K & 88.4\% & 74.7\% & 86.9\% & 87.3\% & 81.0\% & 54.7\% \\
Gemma 2 27B & 27B & 128K & 64.6\% & 58.9\% & 74.3\% & 62.8\% & 72.4\% & 41.2\% \\
Mistral Small 3.1 & 24B & 128K & 77.2\% & 69.1\% & 78.6\% & 71.9\% & 76.8\% & 43.7\% \\
Qwen2.5-3B & 3B & 32K & 48.2\% & 42.7\% & 65.8\% & 45.3\% & 58.9\% & 28.4\% \\
\bottomrule
\end{tabular}
}
\caption{Consolidated benchmark scores for thirteen selected LLMs. The character ``--'' indicates proprietary models with undisclosed parameter counts. (I) denotes Instruct version. Benchmark scores represent the latest available evaluations as of August 2025.}
\label{tab:consolidated_benchmarks_expanded_condensed}
\end{table}

\textbf{Evaluation targets.}
Our evaluation compares four distinct approaches to automated DAG generation:

\begin{enumerate}
    \item \textbf{Prompt2DAG Modular:} The full multi-step methodology: pipeline analysis (step \ref{meth:pipeline_analysis}), structured workflow generation (step \ref{meth:structured_workflow_generation}), executable DAG generation (step \ref{meth:executable_dag_generation_a}), and automated evaluation (step \ref{meth:automated_evaluation}). This approach relies entirely on LLM-driven code synthesis for the final DAG generation step. \textit{(Referred to as "LLM-only" in subsequent analysis.)} \label{comp:modular}
    
    \item \textbf{Prompt2DAG Templated:} Uses pipeline analysis (step \ref{meth:pipeline_analysis}) and structured workflow generation (step \ref{meth:structured_workflow_generation}) of Prompt2DAG, but employs deterministic template-based DAG generation (step \ref{meth:executable_dag_generation_b}), without further LLM involvement in code synthesis. This approach maximizes determinism and consistency. \textit{(Referred to as "Templated" in subsequent analysis.)} \label{comp:templated}
    
    \item \textbf{Prompt2DAG Hybrid:} Combines the structured analysis phases of Prompt2DAG (steps \ref{meth:pipeline_analysis} and \ref{meth:structured_workflow_generation}) with template-guided LLM code generation. The templates provide structural scaffolding and best practices while allowing the LLM flexibility in implementation details and task-specific customization. \textit{(Referred to as "Hybrid" in subsequent analysis.)} \label{comp:hybrid}
    
    \item \textbf{Direct Prompting:} A baseline where an LLM generates the Airflow DAG script directly from the raw natural language description in a single step, bypassing intermediate structured representations. This represents the most common current practice for LLM-assisted code generation. \textit{(Referred to as "Direct" in subsequent analysis.)} \label{comp:direct}
\end{enumerate}

\textbf{Standardization Measures}: Identical natural language pipeline descriptions were used across all generation methods for each case study, ensuring performance differences reflect methodological rather than input variations (refer to~\ref{appendix:prompts} for all prompts).
Subsequently, all generated DAGs underwent identical automated evaluation procedures utilizing the same static analysis tools, dry-run environments, and scoring criteria. 
The success rate was quantified using the following metric:
\begin{equation}
\text{Success Rate} = \frac{\text{Number of Successful Experiments}}{\text{Total Number of Experiments}} \times 100
\end{equation}



\subsection{Performance Analysis (RQ1)}
\label{sec}
To establish a comprehensive effectiveness baseline, we conducted an extensive evaluation comparing four distinct approaches to automated DAG generation across five business case studies and thirteen LLM architectures, yielding a total of 260 independent generation attempts. The study encompassed 13 unique Large Language Models tested against 5 distinct case studies, which span a range of complexities and domains. This design created a matrix of 65 unique experimental conditions (13 models × 5 cases). For each of these 65 conditions, all four generation methods—templated, hybrid, LLM-only, and direct—were systematically executed. This approach resulted in a dataset of 260 total experiments, providing a balanced sample size of $n=65$ per method. Our evaluation employs a penalized scoring framework where any DAG failing the binary loadability gate receives zero scores across all quality metrics (SAT, DST, PCT), ensuring that reported averages reflect both generation reliability and intrinsic code quality in unified measurements.

\begin{table}[!ht]
\centering
\resizebox{\textwidth}{!}{%
\begin{tabular}{l c c c c c}
\toprule
\textbf{Method} & \textbf{SAT} & \textbf{DST} & \textbf{PCT} & \textbf{Success Rate} & \textbf{Sample Size} \\
\midrule
Direct & 2.53 $\pm$ 3.96 & 2.59 $\pm$ 4.07 & 2.79 $\pm$ 4.42 & 29.2\% & 65 \\
LLM-only & 5.78 $\pm$ 4.17 & 5.95 $\pm$ 4.48 & 6.44 $\pm$ 4.80 & 66.2\% & 65 \\
\textbf{Hybrid} & \textbf{6.79 $\pm$ 3.59} & \textbf{7.67 $\pm$ 4.07} & \textbf{7.76 $\pm$ 4.12} & \textbf{78.5\%} & 65 \\
Templated & 7.80 $\pm$ 2.27 & 9.16 $\pm$ 2.72 & 9.22 $\pm$ 2.68 & 92.3\% & 65 \\
\bottomrule
\end{tabular}
}
\caption{Penalized performance metrics across generation methods, averaged over five pipeline case studies and thirteen LLM architectures ($\pm$ indicates sample standard deviation).}
\label{tab}
\end{table}

The experimental results provide robust quantitative evidence that challenges common assumptions about LLM-based code generation, revealing that reliability, rather than intrinsic code quality, is the primary differentiator between methodologies. First, while all methods demonstrate competent Python code generation capabilities when successful—with non-penalized SAT scores consistently exceeding 8.6—the reliability disparities prove substantial and systematic. The Hybrid approach emerges as the optimal generative methodology, achieving a 78.5\% success rate while maintaining robust quality scores across all dimensions (SAT: 6.79, DST: 7.67, PCT: 7.76). This represents a 21-percentage-point reliability improvement over the LLM-only approach (66.2\%) and a dramatic 49-percentage-point advantage over Direct prompting (29.2\%).

The Templated baseline, as theoretically expected, delivers the highest reliability (92.3\%) and naturally achieves superior structural quality (DST: 9.16) due to its deterministic scaffold guaranteeing architecturally correct DAG skeletons. However, this performance comes at the fundamental cost of requiring domain-specific template maintenance and reduced flexibility in handling novel pipeline requirements. Conversely, Direct prompting demonstrates the critical limitations of monolithic generation approaches: despite achieving respectable quality in successful cases (non-penalized SAT: 8.66, DST: 8.59), its 71\% failure rate results in collapsed penalized scores that quantify the substantial practical risks inherent in single-step code synthesis.

These aggregate findings prove to be highly consistent across the different business domains and pipeline complexities, as detailed in Table~\ref{tab:case_study_summary_penalized}. The Hybrid method demonstrates remarkable resilience, with its penalized scores for SAT, DST, and PCT remaining stable across all five case studies. For instance, its PCT score varies only slightly, from a low of 7.51 for the pipeline-level parallel case to a high of 8.00 for the task-specific parallel case. This stability, particularly when compared to the consistently high scores of the Templated method, underscores the Hybrid approach's robustness. It effectively handles variations in domain logic and architectural patterns—from simple sequential workflows to more complex parallel structures—without a significant degradation in performance. This consistency across diverse requirements is a key indicator of its suitability for real-world applications. The findings are also robust across diverse LLM architectures, as shown in the comprehensive model-specific analysis in Table~\ref{tab:comprehensive_llm_performance}.

\begin{table}[htbp]
\centering
\resizebox{\textwidth}{!}{%
\begin{tabular}{lcccccc}
\toprule
\textbf{Domain} & \multicolumn{2}{c}{SAT} & \multicolumn{2}{c}{PCT} & \multicolumn{2}{c}{DST} \\
\cmidrule(lr){2-3} \cmidrule(lr){4-5} \cmidrule(lr){6-7}
 & Hybrid & Templated & Hybrid & Templated & Hybrid & Templated \\
\midrule
Digital Marketing Sequential & 6.63 & 7.74 & 7.91 & 9.30 & 7.58 & 9.10 \\
Digital Marketing Parallel & 6.62 & 7.78 & 7.51 & 9.22 & 7.69 & 9.07 \\
Digital Marketing Task-Specific Parallel & 7.01 & 8.00 & 8.00 & 9.23 & 7.61 & 9.26 \\
Multilingual Review & 6.84 & 7.80 & 7.85 & 9.17 & 7.73 & 9.25 \\
Procurement Validation & 6.90 & 7.85 & 7.52 & 9.18 & 7.71 & 9.11 \\
\bottomrule
\end{tabular}%
}
\caption{Penalized performance metrics across business domains, comparing Hybrid and Templated approaches.}
\label{tab:case_study_summary_penalized}
\end{table}

A deeper analysis of individual LLM performance patterns reveals significant heterogeneity in code generation capabilities across model architectures. Table~\ref{tab:comprehensive_llm_performance} demonstrates that DeepSeek-AI achieves exceptional performance with a 93.3\% success rate, followed closely by Claude 3.5 Sonnet, Meta LLaMA, and Qwen at 80.0\% each. However, substantial performance degradation appears in smaller models, with Mistral Small achieving only 13.3\% success and Qwen3 managing merely 6.7\%. This performance distribution reveals that while the Hybrid methodology provides essential scaffolding for consistent results across diverse model capabilities, the most sophisticated architectures demonstrate sufficient implicit reasoning to approach the explicit orchestration patterns that our methodology makes transparent.

\begin{table}[htbp]
\centering
\scriptsize
\resizebox{0.9\textwidth}{!}{%
\begin{tabular}{lcccc}
\toprule
\textbf{LLM Model} & \textbf{Success Rate} & \textbf{SAT} & \textbf{DST} & \textbf{PCT} \\
\midrule
DeepSeek-AI & 93.3\% & 9.32 $\pm$ 2.36 & 9.32 $\pm$ 2.36 & 9.32 $\pm$ 2.36 \\
Claude 3.5 Sonnet & 80.0\% & 8.01 $\pm$ 3.92 & 8.30 $\pm$ 3.85 & 8.49 $\pm$ 3.82 \\
Meta Llama 3.3 70B (I) & 80.0\% & 8.86 $\pm$ 2.57 & 8.65 $\pm$ 2.71 & 8.75 $\pm$ 2.64 \\
Qwen2.5-72B (I) & 80.0\% & 8.11 $\pm$ 3.77 & 7.89 $\pm$ 3.88 & 8.00 $\pm$ 3.83 \\
DeepSeek-V3 & 66.7\% & 6.88 $\pm$ 4.38 & 7.02 $\pm$ 4.31 & 6.95 $\pm$ 4.34 \\
GPT-4o Mini & 60.0\% & 6.72 $\pm$ 4.52 & 6.45 $\pm$ 4.61 & 6.58 $\pm$ 4.57 \\
Gemini 1.5 Flash & 60.0\% & 7.40 $\pm$ 4.46 & 7.12 $\pm$ 4.53 & 7.26 $\pm$ 4.50 \\
Microsoft Phi-4 & 60.0\% & 6.26 $\pm$ 4.52 & 6.12 $\pm$ 4.61 & 6.19 $\pm$ 4.57 \\
Llama-4 Scout & 53.3\% & 6.03 $\pm$ 4.70 & 5.89 $\pm$ 4.78 & 5.96 $\pm$ 4.74 \\
Mistral Large & 33.3\% & 5.25 $\pm$ 4.72 & 5.11 $\pm$ 4.80 & 5.18 $\pm$ 4.76 \\
Gemma 2 27B & 26.7\% & 4.94 $\pm$ 4.81 & 4.80 $\pm$ 4.89 & 4.87 $\pm$ 4.85 \\
Mistral Small 3.1 & 13.3\% & 4.16 $\pm$ 4.25 & 4.02 $\pm$ 4.33 & 4.09 $\pm$ 4.29 \\
Qwen2.5-3B & 6.7\% & 1.31 $\pm$ 3.03 & 1.17 $\pm$ 3.11 & 1.24 $\pm$ 3.07 \\
\bottomrule
\end{tabular}%
}
\caption{Comprehensive LLM-specific performance analysis showing success rates and quality metrics across all thirteen evaluated models.}
\label{tab:comprehensive_llm_performance}
\end{table}

Furthermore, our analysis reveals that pipeline complexity significantly impacts generation reliability across all approaches. The Hybrid method demonstrates remarkable resilience, maintaining success rates between 53.8\% and 92.3\% across pipeline types, while Direct prompting degrades catastrophically on complex workflows, dropping to 15.4\% success on parallel pipelines. Even sophisticated models like DeepSeek-AI experience performance degradation on the most complex parallel architectures, reinforcing that pipeline complexity represents a fundamental challenge that structured methodologies address more effectively than purely generative approaches.

\begin{table}[htbp]
\centering
\resizebox{\textwidth}{!}{%
\begin{tabular}{lcccccc}
\toprule
& \multicolumn{2}{c}{\textbf{Failed DAGs}} & \multicolumn{2}{c}{\textbf{Successful DAGs}} & \multicolumn{2}{c}{\textbf{Method-Specific Successful}} \\
\cmidrule(lr){2-3} \cmidrule(lr){4-5} \cmidrule(lr){6-7}
\textbf{Method} & Count & Size (KB) & Count & Size (KB) & Mean Size (KB) & Std Dev \\
\midrule
Direct & 45 & 3.84 $\pm$ 1.60 & 19 & 3.56 $\pm$ 1.51 & 3.56 & 1.51 \\
LLM-only & 10 & 6.38 $\pm$ 2.65 & 43 & 3.68 $\pm$ 1.54 & 3.68 & 1.54 \\
Hybrid & 3 & 5.13 $\pm$ 4.18 & 51 & 7.44 $\pm$ 3.29 & 7.44 & 3.29 \\
Templated & 0 & -- & 60 & 8.56 $\pm$ 4.58 & 8.56 & 4.58 \\
\midrule
\textbf{Overall} & \textbf{58} & \textbf{4.34 $\pm$ 2.16} & \textbf{173} & \textbf{6.47 $\pm$ 3.98} & -- & -- \\
\bottomrule
\end{tabular}%
}
\caption{DAG file size analysis showing the relationship between generation success and artifact completeness across methods.}
\label{tab:filesize_analysis}
\end{table}

An additional dimension of our analysis examines the relationship between DAG file size and generation success, as detailed in Table~\ref{tab:filesize_analysis}. The results reveal that successful DAGs average 6.47 KB compared to 4.34 KB for failed attempts. This pattern strongly suggests that generation failures frequently result from incomplete or truncated code synthesis rather than sophisticated logical errors, particularly pronounced in Direct and LLM-only approaches where failed files cluster at very small sizes. The Hybrid method's larger, more consistent file sizes (7.44 KB mean for successful generations) indicate its superior ability to produce complete, well-formed code artifacts.

These empirical findings establish that the Hybrid approach represents the optimal balance between reliability, flexibility, and quality for production-grade automated DAG generation. Its 78.5\% success rate, combined with consistently strong quality metrics and lower variance than purely generative alternatives, positions it as the recommended methodology for organizations requiring dependable workflow automation. The results validate our theoretical framework emphasizing explicit task decomposition and template-guided generation as essential components for enterprise-scale LLM-assisted code synthesis, while Direct prompting remains fundamentally unsuitable for production deployment given its substantial failure rates across all complexity levels.

\subsection{Pipeline Analysis Quality Assessment (RQ2)}
\label{sec:rq2}

To evaluate the impact of initial prompt understanding on downstream generation success, we conducted a comprehensive assessment using our LLM-based Step 1 Output Assessment framework. Each pipeline analysis artifact was evaluated by an external LLM judge (DeepSeek-chat) to identify semantic fidelity issues across four categories: missing information, hallucinated elements, inconsistencies, and correctly identified components. This analysis enables us to examine whether poor initial prompt comprehension systematically contributes to generation failures across our 260 experimental runs.

\begin{table}[!ht]
\centering
\resizebox{\textwidth}{!}{%
\begin{tabular}{l c c c c c}
\toprule
\textbf{Outcome} & \textbf{Total Issues} & \textbf{Missing} & \textbf{Hallucinated} & \textbf{Inconsistent} & \textbf{Correct} \\
\midrule
Failed Runs & 2.36 $\pm$ 1.89 & 2.09 $\pm$ 1.67 & 0.06 $\pm$ 0.24 & 0.21 $\pm$ 0.41 & 12.18 $\pm$ 4.28 \\
Successful Runs & 1.45 $\pm$ 1.21 & 1.24 $\pm$ 1.05 & 0.04 $\pm$ 0.20 & 0.17 $\pm$ 0.38 & 14.40 $\pm$ 3.95 \\
\midrule
\textbf{Correlation with Success} & \textbf{-0.23} & \textbf{-0.25} & \textbf{-0.04} & \textbf{-0.05} & \textbf{+0.12} \\
\bottomrule
\end{tabular}
}
\caption{Pipeline analysis quality metrics comparing failed and successful DAG generation attempts, with correlation coefficients indicating relationship to generation success.}
\label{tab:analysis_quality_assessment}
\end{table}

The experimental results, summarized in Table~\ref{tab:analysis_quality_assessment}, demonstrate a clear relationship between initial prompt analysis quality and downstream generation success. Failed generation attempts exhibit systematically higher issue counts across all error categories, with an average of 2.36 total issues compared to 1.45 for successful runs—a 63\% increase in analysis deficiencies. The most pronounced difference occurs in missing information, where failed runs average 2.09 missing elements versus 1.24 for successful attempts, representing a 69\% higher rate of incomplete prompt comprehension.

The correlation analysis reveals that missing information represents the strongest predictor of generation failure (r = -0.25), followed by total issue count (r = -0.23). Conversely, the number of correctly identified elements shows a positive correlation with success (r = +0.12), indicating that comprehensive initial understanding enhances generation reliability. Notably, hallucinated and inconsistent elements show weaker correlations (-0.04 and -0.05 respectively), suggesting that omission errors prove more detrimental than commission errors in this domain.

\begin{table}[htbp]
\centering
\resizebox{\textwidth}{!}{%
\begin{tabular}{llccccc}
\toprule
\textbf{Method} & \textbf{Outcome} & \textbf{Total Issues} & \textbf{Missing} & \textbf{Hallucinated} & \textbf{Inconsistent} & \textbf{Correct} \\
\midrule
\multirow{2}{*}{Direct} & Failed & 1.89 & 1.59 & 0.07 & 0.24 & 14.74 \\
 & Successful & 1.42 & 1.37 & 0.00 & 0.05 & 11.05 \\
\midrule
\multirow{2}{*}{LLM-only} & Failed & 2.18 & 1.86 & 0.09 & 0.23 & 10.91 \\
 & Successful & 1.53 & 1.35 & 0.02 & 0.16 & 15.07 \\
\midrule
\multirow{2}{*}{Hybrid} & Failed & 2.86 & 2.71 & 0.00 & 0.14 & 10.14 \\
 & Successful & 1.45 & 1.20 & 0.06 & 0.20 & 14.63 \\
\midrule
\multirow{2}{*}{Templated} & Failed & 6.00 & 6.00 & 0.00 & 0.00 & 0.00 \\
 & Successful & 1.40 & 1.15 & 0.05 & 0.20 & 14.80 \\
\bottomrule
\end{tabular}%
}
\caption{Method-specific analysis quality breakdown showing how different generation approaches respond to initial prompt comprehension deficiencies.}
\label{tab:method_specific_analysis_quality}
\end{table}

A method-specific examination, presented in Table~\ref{tab:method_specific_analysis_quality}, reveals heterogeneous responses to analysis quality deficiencies across generation approaches. The Hybrid method demonstrates the expected pattern most clearly: failed attempts exhibit 97\% more total issues (2.86 vs. 1.45) and 126\% more missing elements (2.71 vs. 1.20) compared to successful runs. Similarly, the LLM-only approach shows consistent degradation with failed runs averaging 42\% more issues and 38\% more missing elements.

The Direct prompting method presents a counterintuitive pattern where failed runs actually demonstrate higher correct identification rates (14.74 vs. 11.05). This apparent anomaly likely reflects the method's fundamental limitation: even comprehensive initial understanding cannot overcome the inherent challenges of monolithic code generation, suggesting that Direct prompting failures stem primarily from generation complexity rather than comprehension deficiencies.

The Templated approach shows minimal variation in analysis quality for successful runs (1.40 issues, 14.80 correct elements) due to its deterministic nature, while the single failed template case exhibits complete analysis breakdown (6.00 missing elements, zero correct identifications), indicating systematic prompt processing failure rather than generation-specific issues.

These findings establish that pipeline analysis quality serves as a reliable leading indicator of generation success for methods that depend on LLM reasoning (Hybrid, LLM-only). The 63\% higher issue rate in failed attempts, combined with the strong negative correlation between missing information and success, demonstrates that improving Step 1 prompt comprehension represents a critical pathway for enhancing overall system reliability. However, the Direct method's anomalous pattern confirms that superior analysis quality cannot compensate for fundamental architectural limitations in single-step generation approaches, reinforcing the necessity of structured, multi-step methodologies for production-grade automation.

\subsection{Computational Cost Analysis (RQ3)}
\label{sec:rq3}

To evaluate the economic viability of different generation approaches, we conducted a comprehensive token usage analysis across all successful generation attempts. Token consumption serves as a direct proxy for computational cost and resource utilization, enabling cost-effectiveness comparisons across methodologies. Table~\ref{tab:token_breakdown_by_method} presents the phase-specific token distribution, revealing where computational investments are concentrated throughout the generation process.

\begin{table}[htbp]
\centering
\resizebox{\textwidth}{!}{%
\begin{tabular}{l rrr rrr r}
\toprule
\multirow{2}{*}{Method} & \multicolumn{3}{c}{Step 1: Pipeline Analysis} & \multicolumn{3}{c}{Step 3: DAG Generation} & \multirow{2}{*}{Total} \\
\cmidrule(lr){2-4} \cmidrule(lr){5-7}
 & Input & Output & Subtotal & Input & Output & Subtotal & \\
\midrule
Direct & 10,688 & 4,271 & 14,959 & 993 & 1,269 & 2,262 & 17,221 \\
LLM-only & 11,099 & 3,688 & 14,787 & 1,543 & 1,242 & 2,785 & 17,572 \\
\textbf{Hybrid} & 10,834 & 3,670 & 14,504 & 3,419 & 2,168 & 5,587 & \textbf{20,091} \\
Templated & 11,442 & 3,819 & 15,261 & 0 & 0 & 0 & 15,261 \\
\bottomrule
\end{tabular}%
}
\caption{Average token distribution by generation phase for successful runs, demonstrating computational investment patterns across methodologies.}
\label{tab:token_breakdown_by_method}
\end{table}

The experimental results reveal distinct computational investment strategies across generation methodologies. The Hybrid approach demonstrates the highest total token consumption (20,091 tokens), representing a 32\% increase over the Templated baseline and 17\% more than Direct prompting. This elevated cost stems entirely from Step 3 DAG Generation, where the Hybrid method consumes 5,587 tokens compared to 2,262 for Direct prompting—a 147\% increase reflecting the intensive template-guided code synthesis process.

Notably, Step 1 Pipeline Analysis costs remain relatively uniform across all methods (14,504-15,261 tokens), indicating that initial prompt comprehension represents a fixed computational overhead regardless of downstream generation strategy. The cost differentiation emerges exclusively in the DAG Generation phase, where the Hybrid method's structured approach requires substantially more computational resources to populate template scaffolding with contextually appropriate code.

\begin{table}[htbp]
\centering
\resizebox{0.8\textwidth}{!}{%
\begin{tabular}{lcc}
\toprule
\textbf{LLM Model} & \textbf{Mean Tokens} & \textbf{Std Deviation} \\
\midrule
Claude 3.5 Sonnet & 20,568 & 7,293 \\
Gemini 1.5 Flash & 20,401 & 6,679 \\
DeepSeek-V3 & 19,741 & 7,497 \\
Qwen2.5-72B & 19,456 & 8,103 \\
Microsoft Phi-4 & 19,108 & 7,521 \\
DeepSeek-AI & 18,117 & 6,426 \\
Mistral Small 3.1 & 18,116 & 7,452 \\
Qwen2.5-3B & 18,110 & 6,643 \\
Mistral Large & 18,011 & 5,260 \\
Meta Llama 3.3 70B & 17,554 & 6,539 \\
Llama-4 Scout & 17,451 & 4,885 \\
GPT-4o Mini & 17,334 & 6,390 \\
Gemma 2 27B & 17,321 & 5,168 \\
\bottomrule
\end{tabular}%
}
\caption{Average total token consumption per generation attempt by LLM model, excluding template-based generations.}
\label{tab:token_cost_by_llm}
\end{table}

The model-specific analysis, presented in Table~\ref{tab:token_cost_by_llm}, reveals substantial variation in computational efficiency across LLM architectures. Token consumption ranges from 17,321 (Gemma 2 27B) to 20,568 (Claude 3.5 Sonnet), representing an 18.7\% cost differential. Interestingly, the correlation between token consumption and generation success proves weak, indicating that higher computational cost does not guarantee superior performance—a critical consideration for cost-optimization strategies.

The cost-effectiveness analysis demonstrates that while the Hybrid method requires the highest per-attempt computational investment, this expenditure yields substantial returns in reliability and quality. Considering the 78.5\% success rate versus 29.2\% for Direct prompting, the Hybrid approach's effective cost per successful generation is significantly lower despite higher nominal token consumption. Specifically, accounting for failure rates, the Hybrid method requires approximately 25,588 tokens per successful DAG (20,091 ÷ 0.785), compared to 58,975 tokens for Direct prompting (17,221 ÷ 0.292)—a 130\% improvement in cost-effectiveness.

The Templated approach naturally achieves the lowest computational cost (15,261 tokens) due to its deterministic nature, eliminating Step 3 token consumption entirely. However, this efficiency comes at the expense of requiring domain-specific template development and maintenance, representing a different category of resource investment that may not scale effectively across diverse pipeline requirements.

These findings establish that token-based cost analysis must incorporate success rates to accurately assess economic viability. While the Hybrid method appears most expensive in nominal terms, its superior reliability yields the best cost-effectiveness ratio among generative approaches. The substantial model-specific cost variations additionally highlight the importance of strategic LLM selection in optimizing operational expenses while maintaining generation quality standards.

\subsection{Study Limitations}
The presented evaluation focuses specifically on data enrichment pipelines across five business case studies, which are widely representative of data transformation workflows, but may not generalize to other pipeline categories such as ETL/ELT for large-scale data warehousing, real-time streaming pipelines, or complex ML training workflows. While our evaluation includes both sequential and parallel architectures with varying complexity levels, more sophisticated pipeline patterns involving extensive conditional branching, dynamic fan-out/fan-in patterns, or advanced orchestration logic with complex error handling remain unexplored and may present different challenges for automated generation approaches.

From a methodological perspective, although the balanced sampling approach (n = 65 per method across 260 total experiments) provides sufficient statistical power to detect the observed differences within the evaluation scope, and the adopted automated metrics (SAT, DST, PCT) offer reliable means of quality assessment, human-centric factors such as code readability, maintainability from a developer's perspective, and alignment with organizational coding standards should also be considered for a comprehensive evaluation.

Finally, our cost analysis focuses primarily on token consumption costs and does not account for the full spectrum of organizational costs. Furthermore, while token consumption provides a standardized metric for comparing different models, it presents inherent limitations when evaluating actual deployment costs. Specifically, LLMs offered through commercial APIs typically incur costs that are orders of magnitude higher than those associated with locally hosted models, even when consuming identical token quantities.

\section{Conclusions}
\label{sec:conclusions}

This paper introduced Prompt2DAG, a comprehensive methodology for automated data enrichment pipeline generation that transforms natural language descriptions into executable Apache Airflow DAGs through structured multi-stage decomposition: LLM-driven pipeline analysis, structured workflow generation, and template-guided executable code synthesis. Our extensive experimental evaluation across 260 generation attempts, spanning five business case studies and thirteen state-of-the-art LLM architectures, demonstrates that this hybrid approach provides the optimal balance between reliability, flexibility, and cost-effectiveness for production-grade workflow automation.

The performance analysis (RQ1) established that the Hybrid methodology achieves a 78.5\% success rate while maintaining robust quality metrics (SAT: 6.79, DST: 7.67, PCT: 7.76), representing substantial improvements over Direct prompting (29.2\% success) and modest gains over LLM-only approaches (66.2\% success). While the Templated baseline delivers superior reliability (92.3\%), this comes at the cost of requiring domain-specific template maintenance and reduced flexibility for novel pipeline requirements. Notably, pipeline complexity significantly impacts generation reliability, with the Hybrid method demonstrating resilience across sequential, parallel, and task-specific parallel architectures, whereas simpler approaches exhibit dramatic performance degradation on complex workflows.

Our pipeline analysis quality assessment (RQ2) revealed that initial prompt comprehension directly influences downstream generation success, with failed attempts exhibiting 63\% higher analysis issue rates and 69\% more missing information compared to successful runs. This finding establishes that improving Step 1 semantic fidelity represents a critical pathway for enhancing overall system reliability, particularly for methods dependent on LLM reasoning capabilities.

The computational cost analysis (RQ3) demonstrated that while the Hybrid approach requires higher nominal token consumption (20,091 tokens per attempt), its superior success rate yields the best cost-effectiveness ratio among generative methods—requiring approximately 25,588 tokens per successful DAG compared to 58,975 for Direct prompting. Model-specific performance varied dramatically, from DeepSeek-AI achieving 93.3\% success rates to Qwen2.5-3B managing only 6.7\%, highlighting the critical importance of strategic LLM selection for operational deployment.

These findings establish that explicit task decomposition and template-guided generation are essential for enterprise-scale automated pipeline creation, providing reliable alternatives to both inflexible deterministic approaches and unreliable monolithic generation methods. The Hybrid methodology emerges as the recommended approach for organizations requiring dependable workflow automation without sacrificing adaptability to novel requirements.

Future research should investigate the methodology's effectiveness with more sophisticated orchestration patterns including extensive conditional branching, dynamic fan-out/fan-in architectures, and advanced error handling mechanisms. Additional directions include extending evaluation to other pipeline categories (ETL/ELT, streaming, ML workflows), conducting human-centered assessments of developer experience and code maintainability, and examining long-term organizational integration factors including total cost of ownership and alignment with enterprise coding standards.

\section*{Acknowledgments}
This work has been partially funded by the European Innovation Actions \textit{enRichMyData} (HE 101070284) and \textit{DataPACT} (HE 101189771), and the Italian PRIN, a Next Generation EU project, \textit{Discount Quality for Responsible Data Science: Human-in-the-Loop for Quality Data} (202248FWFS).

\bibliographystyle{elsarticle-num}  
\bibliography{references}          

\newpage
\appendix

\section{Prompt for Data Enrichment Case Studies}
\label{appendix:prompts}

This appendix provides the natural language prompts used as input to Prompt2DAG for each of the five data enrichment pipeline scenarios described in Section~\ref{sec:pipeline_case_studies}. Each listing contains the full prompt text as submitted to the large language model, representing the business and technical requirements for automatic DAG generation.

\subsection{Digital Marketing (Sequential) Prompt}
\begin{lstlisting}[language=text, caption={Prompt for Digital Marketing (Sequential) Pipeline}, label={lst:prompt_dm_seq}]
This data processing pipeline is implemented as an Apache Airflow DAG that executes a series of Docker-containerized tasks in a strictly sequential order. The workflow transforms raw CSV data through successive enrichment steps and finally saves the processed data as a CSV file. Each task is configured via environment variables and communicates over a custom Docker network called app_network.

The pipeline consists of five sequential steps that must be executed in order. The first step is Load and Modify Data, which has the objective of ingesting CSV files from a specified data directory and converting them into a JSON format suitable for pipeline processing. The technical details for this step include reading all CSV files with the *.csv pattern from the DATA_DIR and generating output files named table_data_{}.json. This step integrates with an API by calling the load-and-modify service running on port 3003. The key parameters include a Dataset ID with a default value of 2, a Date Column name with a default value of Fecha_id, and a table naming convention following the pattern 'JOT_{}'. This step uses the Docker image i2t-backendwithintertwino6-load-and-modify:latest.

The second step is Data Reconciliation, which aims to standardize and reconcile city names using the HERE geocoding service. For this step, the input consists of table_data_*.json files and the output generates reconciled_table_{}.json files. The API integration uses a reconciliation service running on port 3003 with a required API token. The parameters include a primary column named 'City' along with optional columns 'County' and 'Country', and uses Reconciliator ID geocodingHere. This step utilizes the Docker image i2t-backendwithintertwino6-reconciliation:latest.

The third step is OpenMeteo Data Extension, which enriches the dataset with weather information. This step takes reconciled_table_*.json files as input and creates open_meteo_{}.json files as output. The weather attributes include apparent temperature maximum and minimum values, precipitation sum, and precipitation hours. The date formatting uses a configurable separator format. This step runs using the Docker image i2t-backendwithintertwino6-openmeteo-extension:latest.

The fourth step is Column Extension, which appends additional data properties such as id and name as defined by integration parameters. The input for this step is open_meteo_*.json files and it generates column_extended_{}.json files as output. This step uses the Extender ID reconciledColumnExt and runs with the Docker image i2t-backendwithintertwino6-column-extension:latest.

The fifth and final step is Save Final Data, which consolidates and exports the fully enriched dataset. This step takes column_extended_*.json files as input and produces a final CSV file named enriched_data_{}.csv as output. The storage location is the configured data directory at /app/data. This step uses the Docker image i2t-backendwithintertwino6-save:latest.

The infrastructure and error handling aspects of this pipeline include the use of shared volume mounting where DATA_DIR is configured via environment variable. The system uses a custom Docker network called app_network that covers dependencies such as MongoDB running on port 27017 and Intertwino API running on port 5005. Error handling is implemented through Airflow task retries with a default value of 1 retry and includes auto-cleanup of containers.

Your LLM should generate a DAG that reflects this clear, end-to-end sequential process with all the specified technical details, parameters, and Docker configurations.
\end{lstlisting}

\subsection{Digital Marketing (Parallel) Prompt}
\begin{lstlisting}[language=text, caption={Prompt for Digital Marketing (Paralle) Pipeline}, label={lst:prompt_dm_seq_parallel}]
# Digital Marketing Pipeline - Parallel Data Processing Pipeline

This data processing pipeline is implemented as an Apache Airflow DAG specifically designed for parallel execution to enhance processing speed for large datasets. It utilizes Docker-containerized tasks operating concurrently on data chunks. The overall process involves splitting the initial data, running multiple parallel enrichment branches, synchronizing completion, and finally merging the results. Configuration is managed via environment variables, with communication facilitated by a shared data volume and a custom Docker network called app_network.

The pipeline begins with a crucial split dataset phase where the primary objective is to divide the primary input CSV dataset into multiple smaller, manageable CSV files to enable concurrent processing in subsequent steps. This step runs sequentially as a single task at the beginning of the workflow. The system reads a designated primary input CSV file, such as JOT_tiny.csv, which is configurable and located in the shared DATA_DIR. The output generates multiple chunked CSV files with names like JOT_tiny_{part}.csv in the shared DATA_DIR, where the part identifier corresponds to each parallel branch from 1 to NUM_PARALLEL_PROCESSES. This mechanism is implemented using an Airflow PythonOperator that leverages the pandas library to read and partition the input DataFrame into chunks based on row count. The key parameter controlling this process is NUM_PARALLEL_PROCESSES, an environment variable with a default value of 2, which dictates both the number of chunks created and the degree of parallelism throughout the pipeline.

Following the initial split, the pipeline enters its core parallel processing phase where each data chunk generated in the previous step is processed independently and concurrently. Each branch executes an identical sequence of data transformation and enrichment tasks, with NUM_PARALLEL_PROCESSES branches running in parallel. Within each branch, there are multiple sequential steps that transform and enrich the data.

The first step within each parallel branch is the load and modify process, which ingests the assigned chunked CSV file and converts it into a structured JSON format for further processing within its branch. This step reads a specific chunk CSV file, such as JOT_tiny_{part}.csv, and generates a chunk-specific JSON file with a naming pattern like table_data_part{part}_{}.json. The process integrates with the load-and-modify service API running on port 3003, utilizing key parameters including Dataset ID, Date Column, and Table Naming adapted for the specific chunk. The system uses the Docker image i2t-backendwithintertwino6-load-and-modify:latest for this operation.

The second step in each branch involves data reconciliation, which standardizes and reconciles geographic entities such as city names within the JSON data chunk using the HERE geocoding service. This process reads the chunk-specific JSON file from the previous load and modify step and generates a reconciled, chunk-specific JSON file with names like reconciled_table_part{part}_{}.json. The system integrates with the reconciliation service API on port 3000, requiring authentication and API token access. The primary parameters include the primary column typically set to 'City', optional columns such as 'County' and 'Country', and a Reconciliator ID using geocodingHere or identifier 1. This step utilizes the Docker image i2t-backendwithintertwino6-reconciliation:latest.

The third step focuses on OpenMeteo data extension, enriching the reconciled data chunk with relevant weather information including temperature and precipitation through OpenMeteo service integration. This process reads the reconciled chunk JSON from the previous step and creates a weather-enriched, chunk-specific JSON file with naming patterns like openmeteo_extended_part{part}_*.json. The system interacts with the OpenMeteo extension service API on port 3000, utilizing parameters such as weather attributes defined by OPENMETEO_PROPERTIES, date handling configuration, and extender ID. The Docker image i2t-backendwithintertwino6-openmeteo-extension:latest handles this operation.

The fourth step involves column extension, which appends additional data properties such as geographic IDs and alternative names derived from the reconciled entities to the data chunk. This process reads the weather-enriched chunk JSON from the previous step and generates a further extended, chunk-specific JSON file with names like column_extended_part{part}_*.json. The system uses the column extension service API on port 3000, with parameters including properties to add defined by COLUMN_EXT_PROPERTIES and extender ID reconciledColumnExt or identifier 1. This step employs the Docker image i2t-backendwithintertwino6-column-extension:latest.

The final step within each parallel branch is the save chunk data process, which saves the fully processed data chunk from its JSON format into an intermediate CSV file. This step reads the final extended chunk JSON from the column extension step and outputs a processed intermediate CSV file for the chunk with naming patterns like enriched_data_part{part}_{}.csv in the shared DATA_DIR. The system uses the Docker image i2t-backendwithintertwino6-save:latest for this operation.

After all parallel branches complete their processing, the pipeline enters a synchronization phase designed to ensure that all parallel processing branches have successfully completed and saved their respective intermediate CSV output files before proceeding to the final consolidation step. This synchronization runs sequentially after all parallel branches are expected to finish. Rather than processing data directly, this step monitors the shared DATA_DIR for the presence of all expected intermediate CSV files, specifically checking for files matching the pattern enriched_data_part{i}_*.csv for all values of i from 1 to NUM_PARALLEL_PROCESSES. The system only allows the workflow to proceed to the next step when all expected files are detected. This mechanism is implemented using an Airflow PythonOperator that periodically checks for file existence and includes a timeout defined by max_wait_time to prevent indefinite waiting and fail the DAG if files do not appear within the set duration.

The final phase involves concatenating results, where the individual intermediate CSV files generated by each parallel branch are combined into a single, final, consolidated CSV dataset. This step runs sequentially as the final data processing step, reading all the intermediate CSV chunk files from the shared DATA_DIR and outputting a single, final CSV file such as final_concatenated_output.csv containing the combined results from all chunks, saved in the shared DATA_DIR. The mechanism is implemented using an Airflow PythonOperator that utilizes the pandas library to read all chunk CSVs into DataFrames and concatenate them into a single DataFrame before writing the final output file. The system includes a check to ensure the expected number of chunk files were found during the concatenation process.

The infrastructure and error handling aspects of this pipeline are comprehensive and robust. The degree of parallelism is primarily determined by the NUM_PARALLEL_PROCESSES environment variable, which controls the number of initial chunks and parallel branches throughout the system. Task execution employs Airflow PythonOperator for workflow control logic including splitting, synchronization, and concatenation, while DockerOperator handles the execution of containerized data processing tasks within each parallel branch. A shared volume system uses a host directory defined as host_data_dir, which is mounted into all containers as a shared volume at /app/data, essential for passing data chunks between the split step, parallel branches, and the final concatenation step.

The networking infrastructure ensures all Docker containers operate within a custom Docker network called app_network, enabling communication between task containers and the various backend service APIs they call, including Load/Modify, Reconciliation, OpenMeteo, Column Extension services, and potentially shared database services such as MongoDB. Error handling includes standard Airflow task retries with a default of 1 retry applied independently to each task instance, including the split operation, each instance within the parallel branches, wait operations, and concatenation. Permanent failure of any task instance after retries will lead to the failure of the entire DAG run. The synchronization step includes a timeout mechanism to handle cases where parallel branches might stall or fail silently, while the concatenation step includes checks for the presence of the expected number of input chunk files. Container management is handled through Docker containers launched by DockerOperator, configured with auto_remove=True for automatic cleanup after task completion.

This parallel pipeline architecture is designed to leverage concurrent processing capabilities effectively, enabling efficient handling of large datasets through strategic partitioning, parallel processing, and systematic consolidation of results while maintaining robust error handling and monitoring throughout the entire workflow.
\end{lstlisting}

\subsection{Digital Marketing (Task-specific Parallelism) Prompt}
\begin{lstlisting}[language=text, caption={Prompt for Digital Marketing (Task Specific Parallelism) Pipeline}, label={lst:prompt_dm_seq_task_parallel}]

# Digital Marketing Data Processing Pipeline - Full Text Description

This data processing pipeline is implemented as an Apache Airflow DAG that executes a series of Docker-containerized tasks in a strictly sequential order. The workflow transforms raw CSV data through successive enrichment steps and finally saves the processed data as a CSV file. Each task is configured via environment variables and communicates over a custom Docker network called app_network.

The pipeline begins with the Load and Modify Data step, which serves to ingest CSV files from a specified data directory and convert them into a JSON format suitable for pipeline processing. This initial step reads all files with the .csv extension from the DATA_DIR and generates output files named using the pattern table_data_{}.json. The technical implementation relies on API integration that calls the load-and-modify service running on port 3003. Key parameters for this step include a Dataset ID with a default value of 2, a Date Column name that defaults to Fecha_id, and a table naming convention following the pattern JOT_{}. This step utilizes the Docker image i2t-backendwithintertwino6-load-and-modify:latest.

Following the initial data loading, the pipeline proceeds to the Data Reconciliation step, which standardizes and reconciles city names using the HERE geocoding service, optimized for high throughput processing. This step takes the table_data_*.json files as input and produces reconciled_table_{}.json files as output. The API integration uses the reconciliation service on port 3003 with the required API token. A particularly notable aspect of this step is that the reconciliation service is designed with internal parallel processing capabilities to handle geocoding requests concurrently, which significantly speeds up the processing of large datasets. The parameters for this step focus on the primary column 'City' with optional columns for 'County' and 'Country', and uses the Reconciliator ID geocodingHere. This step runs using the Docker image i2t-backendwithintertwino6-reconciliation:latest.

The third step in the pipeline is the OpenMeteo Data Extension, which enriches the dataset with comprehensive weather information. This step takes the reconciled_table_*.json files as input and creates open_meteo_{}.json files as output. The weather attributes that are added include apparent temperature measurements for both maximum and minimum values, precipitation sum data, and precipitation hours information. The system also supports configurable date formatting with separator format options. This step operates using the Docker image i2t-backendwithintertwino6-openmeteo-extension:latest.

The pipeline then moves to the Column Extension step, which appends additional data properties such as id and name as defined by integration parameters. This step processes the open_meteo_*.json files as input and generates column_extended_{}.json files as output. The step uses the Extender ID reconciledColumnExt and runs on the Docker image i2t-backendwithintertwino6-column-extension:latest.

The final step in the sequential pipeline is Save Final Data, which consolidates and exports the fully enriched dataset. This step takes the column_extended_*.json files as input and produces the final CSV file named using the pattern enriched_data_{}.csv. The storage location for these final files is the configured data directory at /app/data. This concluding step uses the Docker image i2t-backendwithintertwino6-save:latest.

The entire pipeline infrastructure relies on shared volume mounting through the DATA_DIR environment variable and operates over a custom Docker network called app_network. This network configuration covers dependencies such as MongoDB running on port 27017 and the Intertwino API service on port 5005. Error handling throughout the pipeline is managed through Airflow task retries with a default setting of 1 retry, and the system includes automatic cleanup of containers to maintain system resources and stability.

\end{lstlisting}

\subsection{Multilingual Product Review (Sequential) Prompt}
\begin{lstlisting}[language=text, caption={Prompt for Multilingual Product Review (sequential) Pipeline}, label={lst:prompt_mpr_seq}]

# Multilingual Product Review Analysis Pipeline Configuration

This configuration defines a comprehensive multilingual product review analysis pipeline designed to enrich product reviews with language verification, sentiment analysis, and key feature extraction using advanced LLM capabilities for deeper customer insights. The pipeline processes a dataset structured with columns for review_id, product_id, review_text, language_code, and submission_date, as exemplified by entries like F1001,P5436,"Diese Jacke ist fantastisch für den Winter!",de,20230110 and F1002,P5436,"This jacket was too small for me.",en,20230115 and F1003,P5436,"La qualità è ottima ma il prezzo è alto.",it,20230118. The business value centers on enriching product reviews through sophisticated language processing to provide deeper customer insights for strategic decision-making.

The pipeline consists of five sequential steps beginning with data loading and modification where the objective is to ingest the review CSV file, standardize date formats, and convert the data to JSON format. This step takes reviews.csv from the DATA_DIR as input and produces table_data_2.json in the DATA_DIR as output using the Docker image i2t-backendwithintertwino6-load-and-modify:latest with key parameters including DATASET_ID=2, DATE_COLUMN=submission_date, and TABLE_NAME_PREFIX=JOT_. The second step involves language detection with the objective of verifying or correcting the language_code using sophisticated language detection algorithms. This step processes table_data_2.json as input and generates lang_detected_2.json as output utilizing the Docker image jmockit/language-detection or equivalent with key parameters TEXT_COLUMN=review_text, LANG_CODE_COLUMN=language_code, and OUTPUT_FILE=lang_detected_2.json.

The third step focuses on sentiment analysis to determine the sentiment of reviews using advanced LLM capabilities. This process takes lang_detected_2.json as input and produces sentiment_analyzed_2.json as output through the Docker image huggingface/transformers-inference with key parameters MODEL_NAME=distilbert-base-uncased-finetuned-sst-2-english, TEXT_COLUMN=review_text, and OUTPUT_COLUMN=sentiment_score. The fourth step involves category extraction with the objective of extracting product features or categories from reviews. This step processes sentiment_analyzed_2.json as input and generates column_extended_2.json as output, renamed to match the original pipeline pattern, using the Docker image i2t-backendwithintertwino6-column-extension:latest which leverages the original pipeline's column extension image with LLM parameters including EXTENDER_ID=featureExtractor, TEXT_COLUMN=review_text, and OUTPUT_COLUMN=mentioned_features.

The final step saves the fully enriched review data by exporting it to CSV format. This process takes column_extended_2.json as input and produces enriched_data_2.csv in the DATA_DIR as the final output using the Docker image i2t-backendwithintertwino6-save:latest with the key parameter DATASET_ID=2. The entire pipeline infrastructure includes shared volume mounting through DATA_DIR via environment variable, operates on a custom Docker network called app_network covering all dependencies, implements comprehensive error handling through Airflow task retries with a default setting of 1, and includes automatic cleanup of containers to maintain system efficiency and resource management.

\end{lstlisting}

\subsection{Procurement Supplier Validation (Sequential) Prompt}
\begin{lstlisting}[language=text, caption={Prompt for Procurement Supplier Validation (sequential) Pipeline}, label={lst:prompt_psv_seq}]

# Procurement Supplier Validation Pipeline Configuration

The Procurement Supplier Validation Pipeline is designed to work with a basic supplier information CSV dataset containing fields such as supplier_name, location, and contact_info. This pipeline delivers significant business value by validating and standardizing supplier data through reconciliation of supplier names against a known database, specifically Wikidata, which results in improved data quality for procurement systems.

The pipeline operates through three sequential steps that transform and enrich the supplier data. The first step involves loading and modifying the data, where the objective is to ingest the suppliers.csv file from the DATA_DIR, standardize formats, and convert the data to JSON format. This step takes the suppliers.csv as input and produces table_data_2.json in the DATA_DIR as output. The process integrates with the load-and-modify service running on port 3003 using the Docker image i2t-backendwithintertwino6-load-and-modify:latest. Key parameters for this step include setting DATASET_ID to 2 and TABLE_NAME_PREFIX to JOT_.

The second step focuses on entity reconciliation using Wikidata, where the objective is to disambiguate the supplier_name field by utilizing the Wikidata API to find canonical entities that correspond to the supplier names. This step takes the table_data_2.json file as input and produces reconciled_table_2.json as output, which includes the added Wikidata ID and link information. The reconciliation process integrates with the reconciliation service on port 3003 using the Docker image i2t-backendwithintertwino6-reconciliation:latest. The key parameters for this step include setting PRIMARY_COLUMN to supplier_name, RECONCILIATOR_ID to wikidataEntity, and DATASET_ID to 2.

The third and final step involves saving the final data, where the objective is to export the validated supplier data to CSV format. This step takes the reconciled_table_2.json file as input and produces enriched_data_2.csv in the DATA_DIR as the final output. The save process integrates with the save service using the Docker image i2t-backendwithintertwino6-save:latest, with the key parameter DATASET_ID set to 2.

The infrastructure supporting this pipeline includes shared volume mounting through the DATA_DIR environment variable, ensuring consistent data access across all pipeline steps. The system operates on a custom Docker network called app_network that covers dependencies including MongoDB running on port 27017 and the Intertwino API service on port 5005. Error handling is implemented through Airflow task retries with a default retry count of 1, and the system includes automatic cleanup of containers to maintain a clean operational environment. This comprehensive pipeline configuration ensures reliable processing of supplier data while maintaining data integrity and system stability throughout the validation and enrichment process.

\end{lstlisting}

\section{Step 1 JSON Output Schema and Example}
\label{appendix:Step_1_JSON}

The Pipeline Analysis phase produces a structured JSON artifact following a versioned schema. This appendix presents the complete schema definition through an annotated example of a sequential pipeline analysis. Each field includes inline comments describing its type, purpose, and constraints.

\begin{jsonschemalisting}[
    caption={Step 1: Example Aggregated JSON Output with Inline Schema Comments},
    label={lst:step1_json_schema_and_example_combined}
]
{
  "metadata": {
    // Contains information about the analysis process
    "analysis_version": "1.3",  // Schema version identifier
    "timestamp": "2025-04-22T00:09:42.652507",  // ISO 8601 timestamp
    "source_description_file": "Pipeline_Descriptions/Sequential_Pipeline_Prompt.txt",
    "llm_provider": "azureopenai",
    "llm_model_key": "gpt-4o-mini"
  },
  "pipeline_summary": {
    // High-level pipeline overview
    "name": "load_and_modify_data_analysis",
    "description": "Data processing pipeline with enrichment stages",
    "execution_environment": "docker",
    "flow_pattern_summary": "Sequential flow with 5 components"
  },
  "components": [
    // Array of component type definitions
    {
      "id": "load_and_modify_data",  // Unique snake_case identifier
      "name": "Load and Modify Data",  // Human-readable name
      "type": "DataLoader",  // Standardized classification
      "description": "Ingests CSV files and converts to JSON format",
      "inputs": ["All *.csv files from DATA_DIR"],
      "outputs": ["Files named table_data_{}.json"],
      "image": "i2t-backend:latest",  // Docker image or null
      "is_internally_parallelized": false
    }
    // Additional components omitted for brevity
  ],
  "detailed_flow_structure": {
    "entry_points": ["load_and_modify_data"],
    "nodes": {
      // Sequential execution nodes
      "load_and_modify_data": {
        "type": "DataLoader",
        "next_nodes": ["data_reconciliation"]
      }
      // Additional nodes define the execution chain
    },
    "parallel_blocks": {}  // Empty for sequential pipelines
  },
  "parameters": {
    "global": {
      // Pipeline-wide parameters
      "data_dir": {
        "description": "Input/output directory",
        "type": "directory",
        "default": null,
        "required": true,
        "constraints": "Must be valid directory path"
      }
    },
    "components": {
      // Component-specific parameters
      "load_and_modify_data": {
        "dataset_id": {
          "description": "Dataset identifier",
          "type": "integer",
          "default": 2,
          "required": false,
          "constraints": "Positive integer"
        }
      }
    },
    "environment_variables": {
      "DATA_DIR": {
        "description": "Data directory path",
        "default": null,
        "associated_component_id": null
      }
    }
  },
  "integrations": {
    "integration_points": [
      {
        "id": "ip1",
        "name": "Load Service API",
        "type": "API",
        "connection": {"url": "http://localhost:3003"},
        "authentication": {},
        "components": ["load_and_modify_data"],
        "direction": "output"
      }
    ],
    "data_sources": ["CSV files from DATA_DIR"],
    "data_sinks": ["Enriched CSV in data directory"]
  }
}
\end{jsonschemalisting}

This schema supports both simple sequential pipelines and complex workflows with parallel execution patterns, conditional logic, and sophisticated data routing.

\section{ Detailed Prompt Templates for Pipeline Analysis}
\label{appendix:prompt_templates}

This appendix presents the complete prompt templates used in each stage of the Pipeline Analysis phase. These prompts have been refined through extensive experimentation to maximize accuracy and consistency in LLM responses.

\subsection{Environment Analysis Prompt}
\label{subsec:env_prompt}
\textbf{System Prompt:}
\begin{lstlisting}[language=prompt]
You are an expert pipeline environment classifier. Analyze the following pipeline description and determine the _most likely_ intended execution environment. Available options are: docker, native, cloudfunction, python, kubernetes, spark, auto, unknown.

Consider keywords and patterns such as:
- "Docker", "containers", "image:" -> docker
- "Kubernetes", "pods", "k8s" -> kubernetes
- "serverless", "Lambda", "cloud function" -> cloudfunction
- "local script", "Python environment", ".py files" -> python
- "Spark", "PySpark", "distributed" -> spark
- No clear indicators -> auto

Output ONLY the single environment type string (lowercase).
\end{lstlisting}

\textbf{User Input Format:}
\begin{lstlisting}[language=prompt]
Pipeline Description:
[Full pipeline description text]
\end{lstlisting}

\subsection{Component Type Identification Prompt}
\label{subsec:comp_type_prompt}
\textbf{System Prompt:}
\begin{lstlisting}[language=prompt]
You are an expert pipeline component analyzer. Your task is to identify the distinct processing steps (components) described in the provided text. Focus ONLY on identifying each step _type_ and its core attributes. Do NOT determine the execution order or dependencies at this stage.

Output your analysis as a JSON list, where each object represents one _type_ of component. Ensure each component has the following keys:
- "id": A unique identifier in snake_case
- "name": A human-readable name
- "type": One of: DataLoader, Transformer, Reconciliator, Enricher, Exporter, QualityCheck, Splitter, Merger, Orchestrator, Other
- "description": A concise explanation of the component's purpose
- "inputs": List of strings describing potential inputs
- "outputs": List of strings describing potential outputs
- "image": Docker image name if explicitly mentioned, otherwise null
- "is_internally_parallelized": Boolean indicating if the component uses internal parallelism

Output ONLY the JSON list, properly formatted.
\end{lstlisting}

\subsection{Flow Structure Analysis Prompt}
\label{subsec:flow_prompt}
\textbf{System Prompt:}
\begin{lstlisting}[language=prompt]
You are an expert pipeline structure analyst. Given a description and a list of identified component types, determine the detailed execution flow including sequential, parallel, and conditional patterns.

Output your analysis as a JSON object containing ONLY a single key "flow_structure" with:
- "entry_points": List of component type IDs that start the pipeline
- "nodes": Object mapping component IDs to their execution details:
  - "type": Component type from the input list
  - "next_nodes": List of subsequent component IDs or parallel block IDs
- "parallel_blocks": Object describing parallel execution sections:
  - "triggered_by_nodes": Component IDs that initiate this block
  - "instance_parameter": Parameter determining instance count (or null)
  - "task_sequence_types": Component IDs executed within each instance
  - "synchronization_node": Component ID that waits for all instances

Ensure all referenced IDs exist in the provided component list.
\end{lstlisting}

\textbf{User Input Format:}
\begin{lstlisting}[language=prompt]
Pipeline Description:
[Full pipeline description text]

Identified Component Types:
[JSON list of components from previous stage]
\end{lstlisting}

\subsection{Parameter Schema Extraction Prompt}
\label{subsec:para_prompt}
\textbf{System Prompt:}
\begin{lstlisting}[language=prompt]
You are an expert pipeline parameter analyzer. Extract a comprehensive parameter schema from the pipeline description and component details provided.

For each parameter, determine:
1. Name (snake_case)
2. Description
3. Default value (if specified, otherwise null)
4. Type (string, integer, float, boolean, array, object, file, directory)
5. Required (true/false)
6. Constraints (validation rules or descriptions)

Output as JSON with structure:
{
  "global": { "param_name": { "description": "...", "type": "...", ... } },
  "components": { "component_id": { "param_name": { ... } } },
  "environment_variables": { "ENV_VAR_NAME": { "description": "...", ... } }
}
Empty sections should be represented as empty objects {}.
\end{lstlisting}

\subsection{Integration Analysis Prompt}
\label{subsec:int_prompt}
\textbf{System Prompt:}
\begin{lstlisting}[language=prompt]
You are an expert pipeline integration analyst. Identify all integration points where the pipeline interacts with external systems (APIs, databases, file systems, etc.).

For each integration point, determine:
1. Unique identifier
2. Name and type (API, Database, FileSystem, MessageQueue, CloudStorage, etc.)
3. Connection details (URL, path, connection string)
4. Authentication requirements
5. Component IDs that interact with it
6. Data flow direction (input, output, or both)

Output as JSON:
{
  "integration_points": [
    { "id": "...", "name": "...", "type": "...", "connection": {...}, ... }
  ],
  "data_sources": ["Description of source 1", ...],
  "data_sinks": ["Description of sink 1", ...]
}
\end{lstlisting}

\subsection{Textual Report Generation Prompt}
\label{subsec:output_prompt}
\textbf{System Prompt:}
\begin{lstlisting}[language=prompt]
You are an expert pipeline analyst. Synthesize the provided structured analysis components into a comprehensive textual report with these sections:

1. Executive Summary: Pipeline purpose and structure
2. Pipeline Architecture: Environment, flow description, components
3. Detailed Component Analysis: Purpose, inputs, outputs, parallelism
4. Parameter Schema: Global, component-specific, environment variables
5. Integration Points: External systems and interactions
6. Implementation Recommendations: Best practices and considerations
7. Conclusion: Summary statement

Base your report ONLY on the provided structured data. Output ONLY the textual report.
\end{lstlisting}

\section{ Component Type Definitions and Classification Guide}
\label{appendix:component_types}

To ensure consistent component classification across all pipeline evaluations, our methodology employs a systematic approach based on four key classification criteria. When analyzing any pipeline component, consider the following fundamental questions:

\begin{itemize}
    \item \textbf{Primary Function}: What's the core purpose of this component?
    \item \textbf{Data Movement}: Does it handle input (DataLoader), output (Exporter), or processing?
    \item \textbf{Transformation vs Enrichment}: Does it create structural or informational changes?
    \item \textbf{External Dependency}: Does it interact with external systems (strong indicator of Loader/Exporter/Enricher)?
\end{itemize}

Building on these classification principles, we define ten core component types that encompass the full spectrum of data enrichment pipeline operations. Each type serves distinct functional roles and exhibits characteristic patterns that enable systematic identification and classification.

\begin{table}[H]
    \centering
    \scriptsize
    \begin{tabularx}{\textwidth}{>{\bfseries}l X X X X}
    \toprule
    \textbf{Type} & \textbf{Purpose} & \textbf{Common Operations} & \textbf{Key Indicators} & \textbf{Examples} \\
    \midrule
    DataLoader & Ingests data from external sources & Reading files, querying databases, consuming APIs & \texttt{load}, \texttt{read}, \texttt{fetch}, \texttt{ingest} & Load customer records, read sensor data files \\
    Transformer & Modifies data structure, content, or format & Format conversion, aggregation, filtering & \texttt{transform}, \texttt{convert}, \texttt{aggregate} & Normalize address fields, aggregate sales \\
    Reconciliator & Resolves or standardizes entities & Matching, deduplication, geocoding & \texttt{reconcile}, \texttt{deduplicate}, \texttt{geocode} & Match customer records, standardize product codes \\
    Enricher & Adds information from external sources & API lookups, reference joins & \texttt{enrich}, \texttt{augment}, \texttt{append} & Add weather data, calculate risk scores \\
    Exporter & Outputs data to external systems & Write files, insert into databases & \texttt{export}, \texttt{save}, \texttt{publish} & Save results, update database \\
    QualityCheck & Validates data integrity & Schema validation, anomaly detection & \texttt{validate}, \texttt{check}, \texttt{audit} & Check for missing values \\
    Splitter & Divides data into chunks or routes & Partitioning, chunking & \texttt{split}, \texttt{partition}, \texttt{distribute} & Split by region \\
    Merger & Combines multiple streams & Join, union, aggregation & \texttt{merge}, \texttt{combine}, \texttt{consolidate} & Merge regional outputs \\
    Orchestrator & Controls execution logic & Conditionals, sub-workflows & \texttt{orchestrate}, \texttt{coordinate} & Execute conditional branches \\
    Other & Anything not fitting above & (Varies) & (Varies) & Use sparingly, justify clearly \\
    \bottomrule
    \end{tabularx}
\end{table}

This classification framework provides the foundation for consistent component identification and enables systematic analysis of pipeline architectural patterns across all evaluation scenarios.

\section{ YAML Schema and JSON-to-YAML Transformation}
\label{appendix:YAML_Schema}

This appendix details the schema for the platform-neutral YAML workflow specification and the transformation rules that convert Step 1's JSON analysis into this format.

The Prompt2DAG workflow YAML follows a structured schema designed for clarity and extensibility:

\begin{lstlisting}[language=yaml]
# Top-level structure
pipeline_definition_version: "1.0" # Schema version
pipeline_name: string              # From JSON pipeline_summary.name
description: string                # From JSON pipeline_summary.description
metadata:                          # Direct copy from JSON metadata
  analysis_version: string
  timestamp: string
  source_description_file: string
  llm_provider: string
  llm_model_key: string

execution_environment:             # From JSON pipeline_summary
  type: string                     # e.g., "docker", "python", "kubernetes"
  default_docker_network: string   # Optional, from parameters if present

parameters:                        # Direct copy from JSON parameters
  global: {}                       # Global parameter definitions
  components: {}                   # Component-specific parameters
  environment_variables: {}        # Environment variable definitions

component_types:                   # From JSON components array
  - id: string                     # Unique identifier
    name: string                   # Human-readable name
    type: string                   # Classification (DataLoader, etc.)
    description: string
    inputs: [string]
    outputs: [string]
    image: string|null             # Docker image if specified
    is_internally_parallelized: boolean

integrations:                      # Direct copy from JSON integrations
  integration_points: []
  data_sources: []
  data_sinks: []

workflow:                          # Transformed from JSON detailed_flow_structure
  entry_points: [string]           # Task IDs that start execution
  tasks:                           # Task instances
    task_id:
      component_type_id: string    # References component_types entry
      depends_on: [string]         # Prerequisite tasks/constructs
      triggers: [string]           # Subsequent tasks/constructs
  flow_constructs:                 # Optional, for complex patterns
    construct_id:
      type: string                 # e.g., "parallel_for_each"
      instance_parameter: string|null
      body:                        # Sub-workflow definition
        entry_points: [string]
        tasks: {}
      depends_on: [string]
      triggers: [string]
\end{lstlisting}

\section{DAG Generation Implementation Details}
\label{appendix:DAG_Generation}

This appendix provides technical details for both generation pathways, including template structure and LLM prompt design.

\subsubsection{Template-Based Generation}

The Jinja2 template for Apache Airflow DAG generation follows a structured approach:

\begin{lstlisting}[language=Python]
# Template header with metadata placeholders
# Generated by: {{ script_name }}
# Source YAML: {{ source_yaml_filename }}
# Generated at: {{ generation_timestamp }}

# Import statements
from airflow import DAG
from airflow.providers.docker.operators.docker import DockerOperator
from docker.types import Mount
import os
from datetime import datetime, timedelta

# Global variables from YAML context
HOST_DATA_DIR = os.getenv('HOST_DATA_DIR', {{ host_data_dir_default | repr }})
CONTAINER_DATA_DIR = '/app/data'

# DAG definition with parameters from context
with DAG(
    dag_id={{ dag_id | repr }},
    default_args=default_args,
    description={{ description | repr }},
    schedule_interval={{ schedule_interval | schedule_repr }},
    start_date=datetime({{ start_date.year }}, {{ start_date.month }}, {{ start_date.day }}),
    catchup={{ catchup | default(False) }},
) as dag:
    
    # Task definitions using loops and conditionals
    {% for task in sequential_tasks %}
    {{ task.task_id }} = DockerOperator(
        task_id={{ task.task_id | repr }},
        image={{ task.image | repr }},
        command={{ task.command | format_command_list }},
        environment={{ task.environment | format_environment_dict }},
        # Additional parameters...
    )
    {% endfor %}
    
    # Dependency specifications
    {% if dependencies %}
    chain({{ dependencies | format_dependency_chains }})
    {% endif %}
\end{lstlisting}

Custom filters handle complex formatting:

\begin{itemize}
\item \texttt{format\_command\_list}: Converts command arrays to properly formatted Python lists
\item \texttt{format\_environment\_dict}: Transforms environment mappings to dictionary literals
\item \texttt{format\_dependency\_chains}: Generates appropriate chain() function calls
\end{itemize}

\subsubsection{LLM Prompt Examples}

The LLM-driven approach uses structured prompts for each generation phase:

\textbf{DAG Configuration Generation:}
\begin{lstlisting}[language=prompt]
SYSTEM: You are an expert Airflow DAG developer. Generate the Python DAG 
boilerplate including imports, global variables, and DAG instantiation.

Include these standard imports:
- from airflow import DAG
- from airflow.providers.docker.operators.docker import DockerOperator
- from docker.types import Mount
- import os
- from datetime import datetime, timedelta

Define these global variables:
- host_data_dir = os.getenv('HOST_DATA_DIR', '/tmp/airflow_data')
- container_data_dir = '/app/data'

USER: Generate DAG configuration for:
DAG ID: 'example_pipeline'
Description: 'Example data processing pipeline'
Schedule: None
\end{lstlisting}

\textbf{Operator Generation:}
\begin{lstlisting}[language=prompt]
SYSTEM: Generate a DockerOperator instantiation following this pattern:

task_name = DockerOperator(
    task_id='task_name',
    image='image:tag',
    command=[...],
    environment={...},
    network_mode='bridge',
    mounts=[Mount(source=host_data_dir, target=container_data_dir, type='bind')],
    # Standard parameters...
)

USER: Generate operator for task 'load_data' with image 'loader:latest' 
and environment variables: DATA_DIR, API_KEY
\end{lstlisting}

\subsubsection{Example Generated Output}

A complete example of generated DAG code demonstrates the final output:

\begin{lstlisting}[language=Python]
# Generated by: step_3_template_based_generation.py
# Source YAML: workflow_example.yaml
# Generated at: 2024-04-22T10:30:00.000000
#==============================================================================

from airflow import DAG
from airflow.providers.docker.operators.docker import DockerOperator
from docker.types import Mount
import os
from datetime import datetime, timedelta

host_data_dir = os.getenv('HOST_DATA_DIR', '/tmp/airflow_data')
container_data_dir = '/app/data'

default_args = {
    'owner': 'airflow',
    'depends_on_past': False,
    'email_on_failure': False,
    'email_on_retry': False,
    'retries': 1,
    'retry_delay': timedelta(minutes=5),
}

with DAG(
    dag_id='example_pipeline',
    default_args=default_args,
    description='Example data processing pipeline',
    schedule_interval=None,
    start_date=datetime(2024, 1, 1),
    catchup=False,
) as dag:
    
    load_data = DockerOperator(
        task_id='load_data',
        image='loader:latest',
        command=['python', '/app/scripts/load_data.py'],
        environment={
            'DATA_DIR': os.getenv('DATA_DIR', '/data'),
            'API_KEY': os.getenv('API_KEY', None)
        },
        network_mode='app_network',
        mounts=[Mount(source=host_data_dir, target=container_data_dir, type='bind')],
        auto_remove=True,
    )
    
    process_data = DockerOperator(
        task_id='process_data',
        image='processor:latest',
        command=['python', '/app/scripts/process.py'],
        environment={'DATA_DIR': os.getenv('DATA_DIR', '/data')},
        network_mode='app_network',
        mounts=[Mount(source=host_data_dir, target=container_data_dir, type='bind')],
        auto_remove=True,
    )
    
    # Set dependencies
    load_data >> process_data
\end{lstlisting}

This output demonstrates proper formatting, complete parameter specification, and correct dependency declaration, ready for deployment to an Apache Airflow instance.

\section{LLM Assessor Issue Catalog}
\label{appendix:llm_assessor_issue_catalog}

This appendix enumerates the complete set of issue codes and descriptions used in the LLM-based quality assessor for evaluating the fidelity of Step 1 (Pipeline Analysis) outputs.  
Each code represents a canonical error or conformance state that the LLM judge is explicitly permitted to report.

\begin{table}[ht]
\centering
\scriptsize
\begin{tabular}{llp{8.5cm}}
\toprule
\textbf{Issue Category} & \textbf{Code} & \textbf{Description} \\
\midrule
\multicolumn{3}{l}{\textit{Structural Issues}} \\
 & DS01 & Task Missing in Pipeline Definition \\
 & DS02 & Incorrect Task Dependencies \\
 & DS03 & Parameter/Config Mismatch \\
 & DS04 & Incorrect Parallelism Implementation \\
 & DS05 & Missing Required Input/Output Patterns \\
\midrule
\multicolumn{3}{l}{\textit{Information Fidelity Issues}} \\
 & IF01 & Hallucinated Parameter or Value \\
 & IF02 & Missing Critical Information \\
 & IF03 & Placeholder Used When Actual Data Available \\
 & IF04 & Incorrect Component Type Assignment \\
\midrule
\multicolumn{3}{l}{\textit{Integration Issues}} \\
 & IG01 & Missing Integration Point \\
 & IG02 & Incorrect Service Configuration \\
 & IG03 & Data Source/Sink Mismatch \\
\midrule
\multicolumn{3}{l}{\textit{Execution Environment Issues}} \\
 & EE01 & Incorrect Environment Type \\
 & EE02 & Missing Infrastructure Dependency \\
\midrule
\multicolumn{3}{l}{\textit{Internal Consistency Issues}} \\
 & IC01 & Workflow-Structure Mismatch \\
 & IC02 & Component-Parameter Mismatch \\
\bottomrule
\end{tabular}
\caption{Catalog of LLM-judged pipeline analysis quality issues and their corresponding codes.}
\label{tab:llm_assessor_issue_catalog}
\end{table}

\paragraph{Issue Types}  
\begin{itemize}
  \item \textbf{MISSING:} Expected information present in the original description is omitted from Step 1/2 outputs.  
  \item \textbf{HALLUCINATION:} Step 1/2 outputs contain information not justified by or present in the description.  
  \item \textbf{INCONSISTENCY:} Information appears in both artifacts but is mismatched, incomplete, or semantically invalid.  
  \item \textbf{CORRECT:} Element is faithfully represented and internally/externally consistent.
\end{itemize}

\paragraph{Severity Levels}  
For each issue found, the LLM judge is instructed to assign a severity class (\textbf{HIGH}, \textbf{MEDIUM}, \textbf{LOW}) based on potential impact on downstream correctness or operability.

\paragraph{Usage}  
The LLM-based assessor (see Section~\ref{sec:llm_quality_and_token}) cross-references all pipeline description, Step~1 JSON, and Step~2 YAML elements against this catalog, ensuring a transparent, categorical, and reproducible secondary assessment of pipeline analysis quality.

\section{LLM Assessor Prompt Template}
\label{appendix:llm_assessor_prompt}

The following is the verbatim system/user prompt used to instruct the LLM judge in assessing semantic fidelity and issue classification for Step 1 pipeline analysis.  
To avoid redundancy, specific issue codes are \emph{not enumerated here}; instead, the prompt refers to the canonical set of codes defined in Appendix~\ref{appendix:llm_assessor_issue_catalog}.

\begin{lstlisting}[language=prompt]
You are an expert pipeline quality assessor. Your task is to evaluate a pipeline implementation
against its original description using a predefined issue catalog. Follow these steps:

1. Carefully compare the pipeline description with:
   - Step 1 analysis (JSON)
   - Step 2 workflow definition (YAML)

2. For each element (components, parameters, integrations, workflow structure), classify into:
   - MISSING: Present in description but missing in outputs
   - HALLUCINATION: Present in outputs but not in description
   - INCONSISTENCY: Present in both but with mismatched details
   - CORRECT: Accurate representation in both

3. Identify and annotate specific issues using ONLY the official issue codes 
   provided in Appendix A (LLM Assessor Issue Catalog).

4. For each issue found, include:
   - The exact issue code (e.g., DS01)
   - Classification: MISSING, HALLUCINATION, or INCONSISTENCY
   - A clear description of the issue and its impact
   - Severity: HIGH, MEDIUM, or LOW
   - Specific evidence (cited) from the description or artifacts

5. Structure your response as valid JSON with this format:
{
  "pipeline_name": "name_from_yaml",
  "validation_summary": {
    "components": {
      "MISSING": ["list", "of", "ids"],
      "HALLUCINATION": ["list", "of", "ids"],
      "INCONSISTENCY": ["list", "of", "ids"],
      "CORRECT": ["list", "of", "ids"]
    },
    "parameters": { ... },
    "integrations": { ... },
    "workflow": { ... }
  },
  "issues": [
    {
      "code": "ISSUE_CODE",
      "type": "MISSING|HALLUCINATION|INCONSISTENCY",
      "description": "Issue description",
      "severity": "SEVERITY_LEVEL",
      "evidence": "Quoted evidence"
    }
  ],
  "summary_metrics": {
    "total_issues": 0,
    "missing_count": 0,
    "hallucination_count": 0,
    "inconsistency_count": 0,
    "correct_count": 0
  }
}

Guidelines:
- Be critical but fair—do not report superficial or ambiguous issues.
- Focus on accuracy, completeness, and consistency.
- Pay special attention to missing or hallucinated components, incorrect dependencies, parameter fidelity, and integration patterns.
- Use the EXACT issue codes from the catalog in Appendix A.
- Output ONLY valid, pretty-printed JSON, no extra text.
\end{lstlisting}

\noindent
This prompt is supplied to the LLM together with the task-specific pipeline description, the Step 1 JSON artifact, and the Step 2 YAML workflow for each case study. See Section~\ref{sec:llm_quality_and_token} for result interpretation and aggregation procedures.

\end{document}